\pdfoutput=1
\documentclass[11pt,twoside,a4paper,cmspaper,final,collab]{cms-tdr}
\def\svnVersion{7afbbb6}\def\svnDate{2025/12/03}\def\cmsCernNoTag{CERN-EP-2025-118}\def\cmsCernDate{\today}\def\cmsMessage{Published in Nature as \href{http://dx.doi.org/10.1038/s41586-025-09711-7}{\doi{10.1038/s41586-025-09711-7}}.}
\begin{document}\cmsNoteHeader{BPH-24-002}

\newlength\cmsTabSkip\setlength{\cmsTabSkip}{1ex}

\cmsNoteHeader{BPH-24-002}
\title{Determination of the spin and parity of all-charm tetraquarks}

\date{\today}

\abstract{
The traditional quark model~\cite{Gell-Mann:1964ewy, Zweig:1964jf} accounts for the existence of baryons, such as protons and neutrons, which consist of three quarks, as well as mesons, composed of a quark-antiquark pair. Only recently has substantial evidence started to accumulate for exotic states composed of four or five quarks and antiquarks~\cite{ParticleDataGroup:2024cfk}. The exact nature of their internal structure remains uncertain~\cite{Berezhnoy:2011xn,Wu:2016vtq,Karliner:2016zzc,Richard:2017vry,Anwar:2017toa,Anwar:2017toa,Nielsen:2018ytt,Nielsen:2018uyn,Ali:2019roi,Brambilla:2019esw,Liu:2019zuc,Wang:2019rdo,Bedolla:2019zwg,liu:2020eha,Jin:2020jfc,Becchi:2020uvq,Chen:2020xwe,Wang:2020wrp,Dong:2020nwy,Zhang:2020hoh,Zhu:2020xni,Liu:2021rtn,Faustov:2022mvs,Feng:2023agq,Celiberto:2024mab,Belov:2024qyi,Zhu:2024swp}. This paper reports the first measurement of quantum numbers of the recently discovered family of three all-charm tetraquarks~\cite{LHCb:2020bwg,ATLAS:2023bft,CMS:2023owd}, using data collected by the CMS experiment at the Large Hadron Collider from 2016 to 2018~\cite{CMS:2008xjf,CMS:2023gfb}. The angular analysis techniques developed for the discovery and characterization of the Higgs boson~\cite{Bolognesi:2012mm,CMS:2012qbp,CMS:2012vby} have been applied to the new exotic states. Here we show that the quantum numbers for parity $P$ and charge conjugation $C$ symmetries are found to be $+1$. The spin $J$ of these exotic states is determined to be consistent with $2\hbar$, while $0\hbar$ and $1\hbar$ are excluded at 95\% and 99\% confidence level, respectively. The $J^{PC}=2^{++}$ assignment implies particular configurations of constituent spins and orbital angular momenta, which constrain the possible internal structure of these tetraquarks.
}

\hypersetup{
pdfauthor={CMS Collaboration},
pdftitle={Determination of the spin and parity of all-charm tetraquarks},
pdfsubject={CMS},
pdfkeywords={CMS, tetraquark, exotic hadron, charmonium, spin, parity},
} 

\maketitle 

\section{The exotic states}
\label{sec:nature_introduction}

In 1964, Gell-Mann~\cite{Gell-Mann:1964ewy} and Zweig~\cite{Zweig:1964jf} independently proposed that hadrons, 
such as protons and neutrons, are made up of elementary particles called quarks ($\PQq$).
They suggested that mesons and baryons consisted of quark-antiquark pairs 
($\PQq\PAQq$) and quark triplets ($\PQq\PQq\PQq$), respectively.
At the time, only three quark types were proposed: the up (\PQu) and down (\PQd), 
which form protons ($\PQu\PQu\PQd$) and neutrons ($\PQu\PQd\PQd$), 
and the strange quarks. The discovery of the \PJGy meson in 1974~\cite{E598:1974sol,SLAC-SP-017:1974ind} 
was soon recognized as evidence of a fourth quark type, charm (\PQc), bound with its corresponding antiquark.
Gell-Mann and Zweig also pointed out that the symmetry principles underlying their model allowed 
for ``exotic" configurations, such as tetraquark and pentaquark systems. Despite numerous searches, 
the existence of these particles remained uncertain by the end of the last century~\cite{Landsberg:1999rx}.

The study of exotic hadrons advanced when access to large samples of hadrons containing 
heavy $\PQc$ and $\PQb$  quarks became available, highlighted by the 2003 discovery
of the new particle $\PXP{3872}$ in $\PJGy\,\PGpp \PGpm$ decays~\cite{Belle:2003nnu}.
This particle has been interpreted as a tetraquark candidate with content ${\PQu}{\PQc}{\PAQu}{\PAQc}$ 
and recently classified as $\chi_{\PQc1}(3872)$~\cite{ParticleDataGroup:2024cfk}, where the number 
denotes its mass in MeV (megaelectronvolts) and we adopt natural units by setting $c=1$ and $\hbar=1$.
The discovery of several other candidates for tetraquarks and pentaquarks followed, 
including a ${\PQu}{\PQc}{\PAQd}{\PAQc}$ candidate
$\PZ_{\PQc}(3900)^{+}\to\PJGy\PGpp$~\cite{BESIII:2013ris,Belle:2013yex},
also classified as $\mathrm{T}_{\PQc\PAQc1}(3900)^+$~\cite{ParticleDataGroup:2024cfk}.
However, the true nature of their structure continues to be widely 
debated~\cite{Berezhnoy:2011xn,
Wu:2016vtq,Karliner:2016zzc,
Richard:2017vry,Anwar:2017toa,Anwar:2017toa,
Nielsen:2018ytt,Nielsen:2018uyn,Ali:2019roi,Brambilla:2019esw,Liu:2019zuc,Wang:2019rdo,Bedolla:2019zwg,
liu:2020eha,Jin:2020jfc,Becchi:2020uvq,Chen:2020xwe,Wang:2020wrp,Dong:2020nwy,
Zhang:2020hoh,Zhu:2020xni,Liu:2021rtn,Faustov:2022mvs,
Feng:2023agq,Celiberto:2024mab,Belov:2024qyi,Zhu:2024swp}.
Most interpretations fall into two categories: tightly-bound states of two quarks 
and two antiquarks, similar to how quarks are bound within protons and neutrons, or loosely-bound molecules 
composed of two mesons, similar to how protons and neutrons compose an atomic nucleus.

\begin{figure*}[t]
  \centering
    \includegraphics[width=0.9\linewidth]{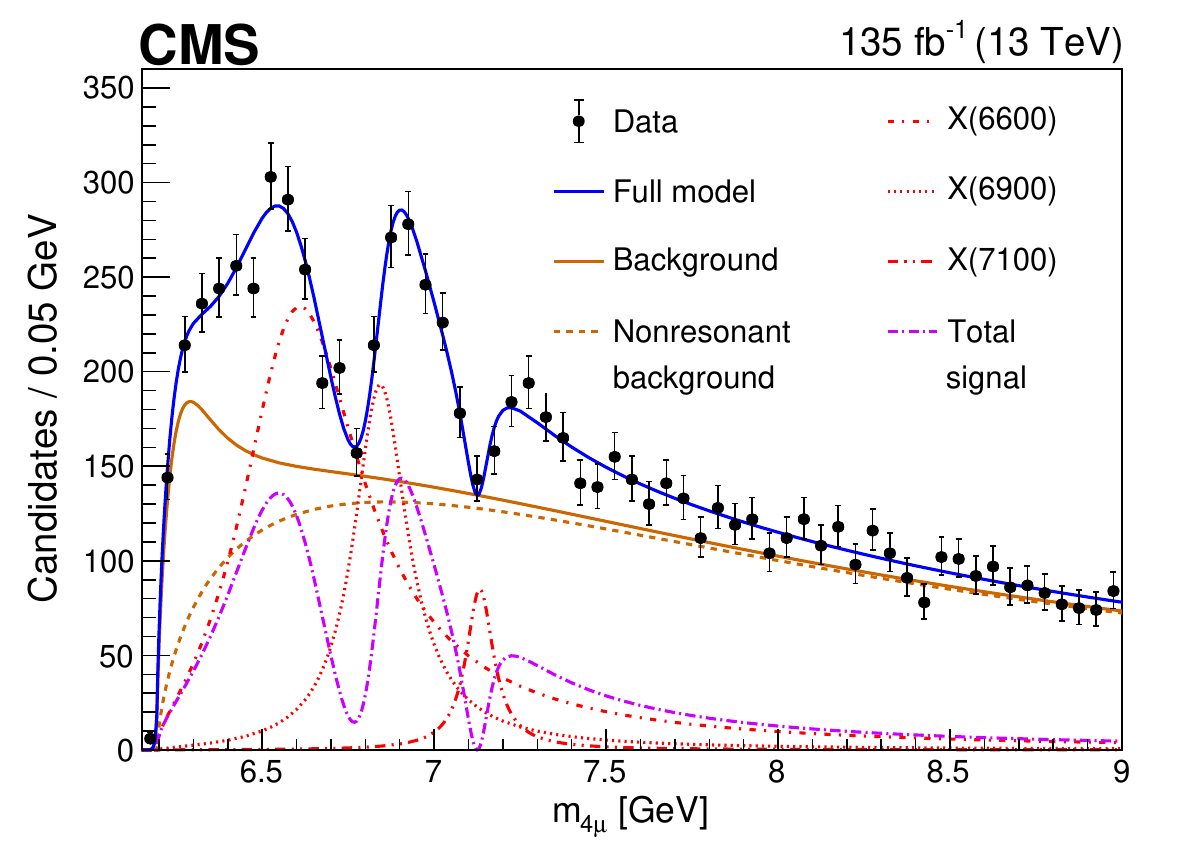}
    \topcaption{
\textbf{Candidates for all-charm tetraquarks.}
The $\PJGy\,\PJGy\to\PGmp\PGmm\PGmp\PGmm$ invariant mass $m_{4\PGm}$ spectrum shows the three exotic states, $\PXP{6600}$, 
$\PXP{6900}$, and $\PXP{7100}$. Parameterizations of these states are displayed both individually and as a combined 
signal that includes quantum-mechanical interference (denoted by ``Total signal"). 
The full model~\cite{CMS:2023owd} incorporates both signal and background components, with the background originating 
from di-\PJGy production, including contributions from nonresonant production and an enhancement 
near the kinematic threshold of 6.2\GeV. 
    }
    \label{fig:mass}
\end{figure*}

The next breakthrough in understanding exotic hadrons may come from studying those composed entirely of heavy quarks.
The LHCb, ATLAS, and CMS Collaborations observed a state $\PXP{6900}$~\cite{LHCb:2020bwg,ATLAS:2023bft,CMS:2023owd},
also referred to as $\mathrm{T}_{\PQc\PQc\PAQc\PAQc}(6900)^{0}$~\cite{ParticleDataGroup:2024cfk},
in the $\PJGy\,\PJGy$ final state.
The CMS Collaboration also reported two additional states, $\PXP{6600}$ and $\PXP{7100}$.
The triple-peaking structure in the di-$\PJGy$ invariant mass distribution observed by CMS~\cite{CMS:2023owd}
is displayed in Fig.~\ref{fig:mass} and is best described by a quantum-mechanical interference 
of three amplitudes representing ${\PQc}{\PQc}{\PAQc}{\PAQc}$ states.
The interference pattern among the three resonances, along with the mass spacings that follow a radial Regge 
trajectory~\cite{Zhu:2024swp}, suggests that these three \PX particles form a family of states with identical properties, 
differing only in the radial excitations of their wavefunctions.
These properties, which include the spin and symmetry quantum numbers, provide insights into the internal structure 
of these exotic hadrons and are the primary focus of the experimental study reported in this article.
The corresponding tabulated results are available in the HEPData record for this analysis~\cite{hepdata}.

\section{The spin and symmetry properties}
\label{sec:nature_spin}

Spin is a type of intrinsic angular momentum carried by particles. For composite particles,
the total spin $J$ represents the system's total angular momentum, determined by the combination 
of the spins $S$ and orbital angular momenta $L$ of the elementary constituent particles. 
The quantum number for parity ($P$) symmetry describes how a system behaves under 
a spatial inversion or, equivalently, a mirror reflection. The charge conjugation ($C$) symmetry 
describes how a system transforms when every particle is replaced by its corresponding antiparticle.
The $P=\pm1$ and $C=\pm1$ values indicate whether the wavefunction of the state changes sign under 
the respective transformation.

\begin{figure*}[t]
    \centering
    \includegraphics[width=1\linewidth]{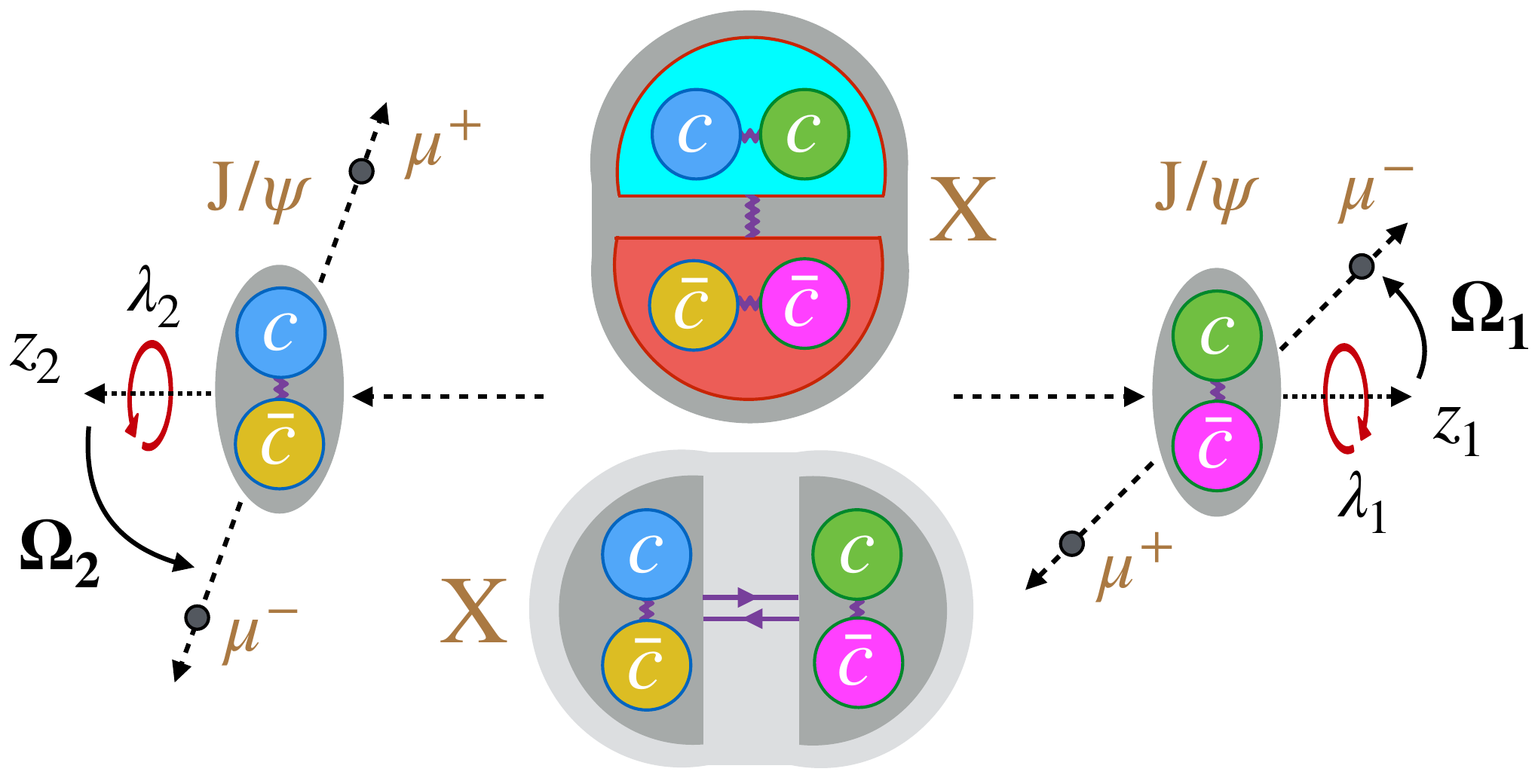}
    \topcaption{
\textbf{Internal structure models for the particle $\PX$.}
The particle $\PX$, composed of ${\PQc}{\PQc}{\PAQc}{\PAQc}$, is shown at rest.
Two models of the internal structure of $\PX$ are presented: 
a tightly-bound tetraquark (upper) and a loosely-bound molecule of two mesons (lower). 
The colours assigned to individual quarks or quark pairs denote possible colour charge assignments in strong interactions, 
where attractive forces are mediated by gluon exchange (shown as wavy lines) and meson exchange (shown as a solid pair of arrows).
The $\PX$ decays into two $\PJGy$ mesons with spin projections $\lambda_i$ along their respective directions of motion; 
each meson then decays into a $\PGmp\PGmm$ pair.
The polar and azimuthal angles $\mathbf{\Omega}_{i}=(\theta_{i}, \Phi_{i})$ describe the direction of the $\PGmm$ 
relative to the $z_i$ axis, which is defined to point opposite to the $\PX$ direction in the centre-of-mass frame 
of the corresponding $\PJGy$ meson, for $i = 1$ and $2$.
}
\label{fig:decay}
\end{figure*}

The quarks and antiquarks have a spin of $S=1/2$, the fundamental quantum of spin,
and they serve as the building blocks of the tetraquark state illustrated in Fig.~\ref{fig:decay}.
The quarks inside a hadron can carry three different colour charges, associated with the strong interaction, 
conventionally referred to as blue, green, and red. 
Antiquarks carry corresponding anticolour charges (antiblue, antigreen, and antired), 
which are depicted as yellow, magenta, and cyan in Fig.~\ref{fig:decay}.
Quarks and antiquarks are held together through the exchange of gluons, 
which are the mediators of the strong interaction.
The hadron as a whole is colour-charge-neutral.

A tightly-bound tetraquark state \PX is illustrated in Fig.~\ref{fig:decay} (upper). 
In the ground state with zero orbital angular momentum, two identical charm quarks form an antisymmetric colour 
state for attraction,  leading to symmetric spatial and spin states with total spin $1$~\cite{Faustov:2022mvs}.
The two charm antiquarks do the same.
Each pair carries a colour charge as well, and the two pairs attract each other, forming a strongly-bound 
tetraquark state that is colour-charge-neutral, similar to a quark-antiquark bound state in a meson.
The orbital angular momentum $L$ between the quark pair and the antiquark pair can take non-negative integer values. 
The corresponding parity of the system is then given by $P=(-1)^{L}$,
which arises from the behavior of the spherical harmonics under spatial inversion.

The lowest and most probable energy state with $L=0$ is spatially symmetric, $P=+1$. 
The spins of the two systems combine in a symmetric configuration to yield a total spin $J=0$ or $2$ with $C=+1$,
or in an antisymmetric configuration to give $J = 1$ with $C = -1$.
For the $\PQc\PQc$ and $\PAQc\PAQc$ system, charge conjugation is equivalent to exchanging the pairs, 
resulting in the associated symmetry.
As we will demonstrate later, the $C = -1$ configuration is ruled out and will therefore not be considered further.
For an antisymmetric spatial state with $L=1$ and $P=-1$, the spins combine to $1$ in an antisymmetric configuration
and with $C=+1$, resulting in possible total spins $J=0$, $1$, or $2$.
States with $L = 2$ are also possible, resulting in $P = +1$ and allowing $J$ values up to $4$ when the spins combine to $0$ or $2$.
However, the high orbital angular momentum requires additional energy, making such states less likely.

An alternative model, shown in Fig.~\ref{fig:decay} (lower), is a loosely-bound molecule of two $\PQc\PAQc$ mesons. 
The lowest-energy configuration corresponds to an orbital angular momentum $L=0$ between the two mesons,
resulting in $P=+1$. A key distinction is that, unlike in a tightly-bound tetraquark, the two constituent $\PQc\PAQc$ 
mesons are not restricted to form spin-1 states. Consequently, lower total spin values such as $J=0$ or $1$
are more likely, although higher spin states cannot be excluded.
Another difference is the weaker interaction between the $\PQc\PAQc$ mesons. 
Similar to how a deuteron is a bound state of a proton and a neutron, the two colour-charge-neutral systems 
are bound through the exchange of a meson via the Yukawa interaction~\cite{Yukawa:1935xg}.
However, unlike the deuteron, in an all-charm tetraquark molecule, the exchanged meson must contain charm quarks. 
A heavier exchange meson significantly suppresses the Yukawa interaction, making the formation of bound 
states less likely. However, alternative empirical models for these interactions have also been 
explored~\cite{Brambilla:2019esw,Nielsen:2018ytt,Nielsen:2018uyn,Ali:2019roi}.

The three $\PX$ states under investigation have invariant masses ranging between $6.2$ and $8.0\GeV$ (gigaelectronvolts),
as shown in Fig.~\ref{fig:mass}, mean lifetimes between $10^{-24}$ and $10^{-23}$ seconds~\cite{CMS:2023owd}, 
and they decay into either two $\PJGy$ mesons, or potentially several other, yet unobserved, final states.
The $\PJGy$ meson has a mass of $3.1\GeV$, spin $1$, a mean lifetime of approximately $7 \times 10^{-21}$~seconds,
and in 6\% of the cases, the $\PQc\PAQc$ pair in the $\PJGy$ annihilates into a $\PGmp\PGmm$ pair~\cite{ParticleDataGroup:2024cfk}, 
which is ideal for detection of the final states and for performing an angular analysis.

\section{Angular distributions}
\label{sec:nature_angles}

The $\PX$ quantum numbers can be inferred from the polarizations of the $\PJGy$ mesons, which, 
in turn, result in specific angular distributions of the decay muons. 
Figure~\ref{fig:decay} defines $\lambda_1$ and $\lambda_2$ as the spin projections
of the two $\PJGy$ mesons along their respective directions of motion. 
Both $\lambda_1$ and $\lambda_2$ can take values of $-1$, $0$, or $+1$.
Let $A_{\lambda_1\lambda_2}$ be the quantum mechanical amplitude that describes the decay of 
$\PX$ into two mesons with spin projections $\lambda_1$ and $\lambda_2$.
The nine values of $\abs{A_{\lambda_1\lambda_2}}^2$ can be interpreted as relative probabilities 
for specific polarization states of the two $\PJGy$ mesons. 

The angular distributions in the decay $\PX\to \PJGy\,\PJGy\to (\PGmp\PGmm)( \PGmp\PGmm)$
can be expressed as a function of three angular observables
$\theta_1$, $\theta_2$, and $\Phi=(\Phi_1+\Phi_2)$ as illustrated in Fig.~\ref{fig:decay}
and further explained in Appendix~\ref{sec:nature_append_angles}. 
Here, $\theta_1$ and $\theta_2$ are the helicity angles of the two $\PJGy$ mesons,
and $\Phi$ represents the angle between their decay planes.
The nine amplitudes $A_{\lambda_1\lambda_2}$ appear in the coefficients of the expression.
The analytical and numerical calculations of the angular distributions are detailed in 
Refs.~\cite{Gao:2010qx,Bolognesi:2012mm}. 

We base the prediction of $A_{\lambda_1 \lambda_2}$ on symmetry considerations~\cite{Gao:2010qx}.
Angular momentum conservation implies that $\abs{\lambda_1-\lambda_2} \le J$.
For two identical $\PJGy$ particles, the relation $A_{\lambda_1\lambda_2}=(-1)^{J}A_{\lambda_2\lambda_1}$ must hold.
Finally, assuming that the $\PX$ has definite $P$ and $C$ quantum numbers, which are conserved in the strong decays, 
we conclude that $A_{\lambda_1\lambda_2} = P(-1)^{J} A_{(-\lambda_2)(-\lambda_1)}$ and the charge conjugation $C = +1$.
The latter is due to the $C = +1$ of the $\PJGy\,\PJGy$ final state.
This results in six possible assignments for the $J^{PC}$ quantum numbers with the corresponding contributions 
from the amplitudes shown in Table~\ref{table:jcp}. 
States $J^{P+}$ with $J \ge 3$ would exhibit amplitudes and angular distributions 
similar to those of $2^{P+}$, as inferred from Ref.~\cite{Gao:2010qx}.

\renewcommand{\arraystretch}{1.3}
\begin{table*}[t!]
  \centering
      \topcaption{
      \textbf{Quantum numbers.} 
      The possible assignments of quantum numbers $J^{PC}$, the $J_i^{P}$ models considered, and the contributing amplitudes 
      in the decay $\PX\to \PJGy\,\PJGy$ are presented.
      }
      \begin{tabular}{lll}
      \hline
        $J^{PC}$ & Models $J_i^{P}$ & Contributing amplitudes \\
      \hline
      $0^{-+}$ & $0^{-}$ & $A_{++}=-A_{--}$  \\
      \multirow{2}{*}{$0^{++}$} & $0_m^{+}$  & $A_{00}$  and $A_{++}=A_{--}$  \\
                                              &  $0_h^{+}$ & $A_{++}=A_{--}$  and $A_{00}$   \\
      $1^{-+}$ & $1^{-}$ & $A_{+0}=-A_{0+}=A_{-0}=-A_{0-}$  \\
      $1^{++}$ & $1^{+}$ & $A_{+0}=-A_{0+}=-A_{-0}=A_{0-}$  \\
      \multirow{2}{*}{$2^{-+}$} & $2_m^{-}$  & $A_{++}=-A_{--}$    \\
                                             &  $2_h^{-}$ &  $A_{+0}=A_{0+}=-A_{-0}=-A_{0-}$  \\
      $2^{++}$ & $2_m^{+}$  & $A_{+0}=A_{0+}=A_{-0}=A_{0-}$, $A_{+-}=A_{-+}$, $A_{00}$,  and $A_{++}=A_{--}$  \\
      \hline
      \end{tabular}
      \label{table:jcp}
  \end{table*}

As shown in Table~\ref{table:jcp}, in the scenarios of $0^{-+}$, $1^{-+}$, and $1^{++}$, the relative contributions 
of all amplitudes $A_{\lambda_1 \lambda_2}$ are predetermined. 
In the scenarios of $0^{++}$ and $2^{-+}$, the relative contributions of the two different sets of amplitudes, 
associated with different models $J_i^{P}$ listed in Table~\ref{table:jcp}, are not known a priori. 
The $2^{++}$ scenario has four contributions with an undetermined relationship.
When pure symmetry alone does not provide a determination, the dynamical properties of the interactions 
provide additional constraints that relate the contributions.
Assuming the resonances differ only by radial excitations of their wavefunctions, we expect all three to exhibit 
the same structure in their decay interactions. This structure and the corresponding models 
are defined and motivated in Appendix~\ref{sec:nature_append_model}.

The $0_m^{+}$ model describes the simplest possible structure for the interaction of a spin-0 particle 
decaying into two spin-1 particles, with the subscript $m$ indicating its minimal nature. In contrast, the $0_h^{+}$
model corresponds to a more complex interaction structure, as denoted by the subscript $h$ for higher 
complexity~\cite{Bolognesi:2012mm}.
The $2_m^{-}$ and $2_h^{-}$ models correspond to $A_{++}=-A_{--}$ and $A_{+0}=A_{0+}=-A_{-0}=-A_{0-}$
and their decay angular distributions are indistinguishable from those of $0^{-}$ and $1^{-}$ models, respectively. 
Quantum-mechanical interference between models $0_m^{+}$ and $0_h^{+}$, or $2_m^{-}$ and $2_h^{-}$,  
is also taken into account in the data analysis. The $2_m^{+}$ model, as defined in Ref.~\cite{Bolognesi:2012mm}, 
corresponds to the minimal structure of interactions of a spin-2 particle in its decay. 
It is used to represent the $2^{++}$ quantum numbers with a distinct contribution from $A_{+-} = A_{-+}$, 
which is unique to $2^{++}$. 
Additionally, seven other amplitudes contribute simultaneously, another feature unique to $2^{++}$.
Other possible $2^{++}$ models, beyond the minimal structure, may exhibit angular distributions
similar to those of the $0^{++}$ or $1^{++}$ states, and are not tested separately.

The reasoning outlined above leads to the predicted probability distributions $\mathcal{P}_i(\theta_1, \theta_2, \Phi$, $m_{4\PGm})$ 
for the eight individual $J^{P}_i$ models listed in Table~\ref{table:jcp}, as well as their possible mixtures, 
at any mass value $m_{4\PGm}$ for the particle decay $\PX \to \PJGy\,\PJGy$. 
The aim of the analysis is to identify the model that best matches the experimental data.

\section{Experimental data}
\label{sec:nature_cms}

To create new particles, proton beams are accelerated to a combined energy of 13\TeV (teraelectronvolts)
in opposite directions within the Large Hadron Collider (LHC)~\cite{Evans:2008zzb}, 
an international project hosted by CERN, and collided head-on.
Protons are baryons with a mass of 0.94\,\GeV, composed of three valence quarks.
However, approximately half of their momentum is carried by gluons, which bind the quarks together.
Additionally, sea quark-antiquark pairs emerge and carry roughly one fifth of the proton's momentum.
As a result, new particles can be created through collisions between gluons, quarks, and antiquarks 
(collectively referred to as partons) within the protons.
The data presented in this paper were collected from 2016 to 2018, with a total integrated luminosity 
of $135\fbinv$ (inverse femtobarns), which represents approximately $1.5 \times 10^{16}$ proton-proton ($\Pp\Pp$) collisions.

Once produced, the \PX particles immediately decay within the CMS detector~\cite{CMS:2008xjf,CMS:2023gfb}, a large general-purpose 
international experiment at the LHC\@. Its essential component is a superconducting solenoid with an internal diameter of 6 meters, 
capable of generating a magnetic field of 3.8~tesla. 
Within the solenoid's volume, there are a silicon pixel and strip tracker, 
a lead tungstate crystal electromagnetic calorimeter, and a brass and scintillator hadron calorimeter, each consisting 
of a barrel and two endcap sections. Muons are identified using gas-ionization chambers embedded in the steel 
flux-return yoke outside the solenoid. In this analysis, the primary measurements of muon momentum vectors are 
obtained from the curvature and direction of their paths in the silicon tracker. A fast, real-time data recording decision 
is made using a two-level trigger system~\cite{Khachatryan:2016bia}, which reduces the recorded data rate 
to approximately 1\unit{kilohertz}.

To select events of interest, the trigger criteria include having two or three muons each with a minimum
muon momentum transverse to the beam (\pt) of 3\,\GeV.
The procedure for the offline selection of $\PX$ candidates is described in Ref.~\cite{CMS:2023owd}:
there must be at least four muons, each with $\pt {\,>\,} 2\GeV$;
$\PGmp\PGmm$ pairs must originate from a common vertex, 
have a mass consistent with a $\PJGy$, and $\pt {\,>\,} 3.5\GeV$. 
A fit is applied to muon pairs, using the $\PJGy$ meson mass and the common production vertex as constraints,
in order to improve the $m_{4\PGm}$ resolution. 
A total of $8651$ candidate events with two $\PJGy$ mesons have been recorded 
with a combined mass in the range between 6.2--9.0\,\GeV, 
as shown in Fig.~\ref{fig:mass}, with approximately a quarter expected to correspond to the three \PX states, 
and the remainder mostly originating from background $\PJGy$ pair production.
These candidates are the same as those in Ref.~\cite{CMS:2023owd}.
For each \PX candidate, we record the momentum vectors of all four muons for the analysis that follows.

\section{Angular analysis}
\label{sec:nature_analysis}

The objective of the analysis is to compare the angular distributions in the decay of \PX particles observed in the experiment 
with the predicted distributions for various $J^{PC}$ models and select the model that is consistent with the data.
To prevent a potential bias, the observed distributions were hidden until all aspects of the analysis were finalized.
The predicted distributions $\mathcal{P}_i(\theta_1, \theta_2, \Phi, m_{4\PGm})$ are precisely known for the muons originating 
from the decay, but the observed distributions are modified by detector effects. This is because the detection probabilities 
of the reconstructed muons depend on their momenta and directions, which have complex correlations with the angles.
Additionally, the momenta and directions themselves are subject to uncertainties.
To account for these effects and to accurately model the background, a comprehensive computer simulation 
of  $\Pp\Pp$ collisions is performed using Monte Carlo (MC) techniques. The simulation proceeds in four stages.

In the first stage, the \textsc{JHUGen} program, version 7.5.7~\cite{Gao:2010qx,Bolognesi:2012mm}, originally designed to model 
variations in the Higgs boson's properties, is employed to simulate the process $\Pp\Pp\to \PX \to \PJGy\,\PJGy \to \PGmp\PGmm\PGmp\PGmm$. 
Two models of \PX production are explored~\cite{Brambilla:2019esw, Nielsen:2018ytt,Nielsen:2018uyn,Ali:2019roi}: 
direct production through parton annihilation within protons, and fragmentation of a secondary gluon or quark generated in collisions. 
In the case of $J\ge1$, the first model can lead to a preferred spin polarization of \PX along the beam axis,
while the second model may result in polarization along the direction of the \PX particle's motion.
\textsc{JHUGen} enables the modelling of either polarization, or the simulation of an unpolarized state,
through the application of event weights. Polarization variations are used to estimate uncertainties, 
with the unpolarized case assumed by default.

The three stages of the simulation that follow are shared with thousands of other processes analysed by the 
CMS Collaboration and have therefore been extensively tuned to ensure a good agreement with the experimental data.
The simulation of the remaining particles appearing in proton collisions and surrounding the four muons is performed
by the \PYTHIA program, version 8.240~\cite{Sjostrand:2014zea}. 
The simulation takes into account the effects of extra $\Pp\Pp$ collisions that happen at the same time 
or close in time to the main collision.
The generated events are further processed through a dedicated CMS detector simulation, 
based on the \GEANTfour program~\cite{GEANT4:2002zbu}, which models the detector response.
Finally, the simulated events undergo the complete detector reconstruction process,
using software and algorithms identical to the one used to analyse the experimental data.

The nonresonant $\PJGy\,\PJGy$ background is simulated with \PYTHIA in the first stage and covers cases where the 
two mesons are produced either in the same or in different parton collisions in a single  $\Pp\Pp$ interaction.
The simulated angular distributions were found to be compatible with data in the mass sideband above 8\,\GeV.
The momentum of the $\PJGy\,\PJGy$ pair in all three directions was tuned for both \PX and nonresonant production 
using \PYTHIA settings in the second stage to match the data observed in both the sideband and \PX signal regions.

\begin{figure*}[t!]
    \centering
    \centerline{
      \includegraphics[width=0.46\linewidth]{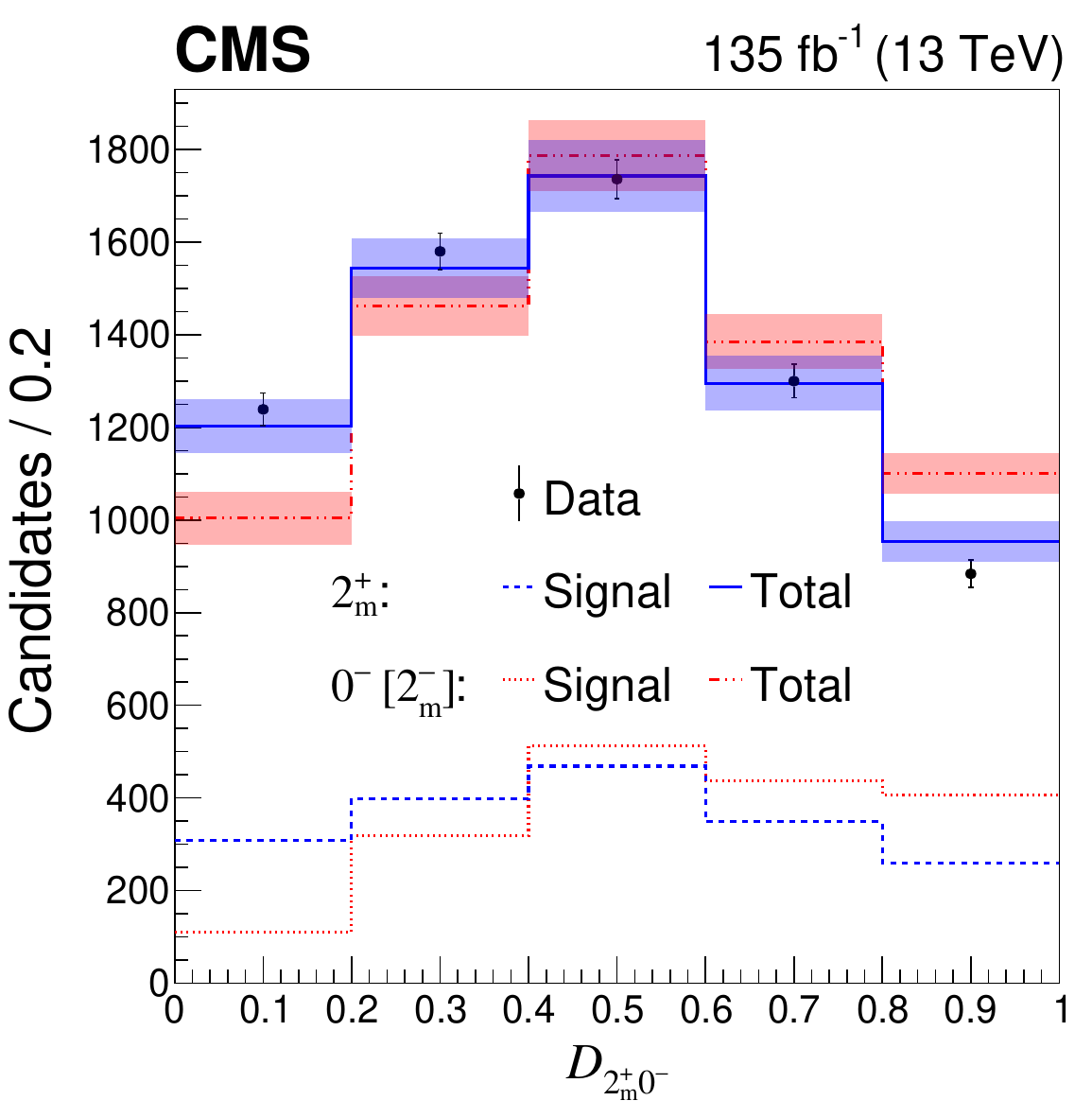}
      \includegraphics[width=0.49\linewidth]{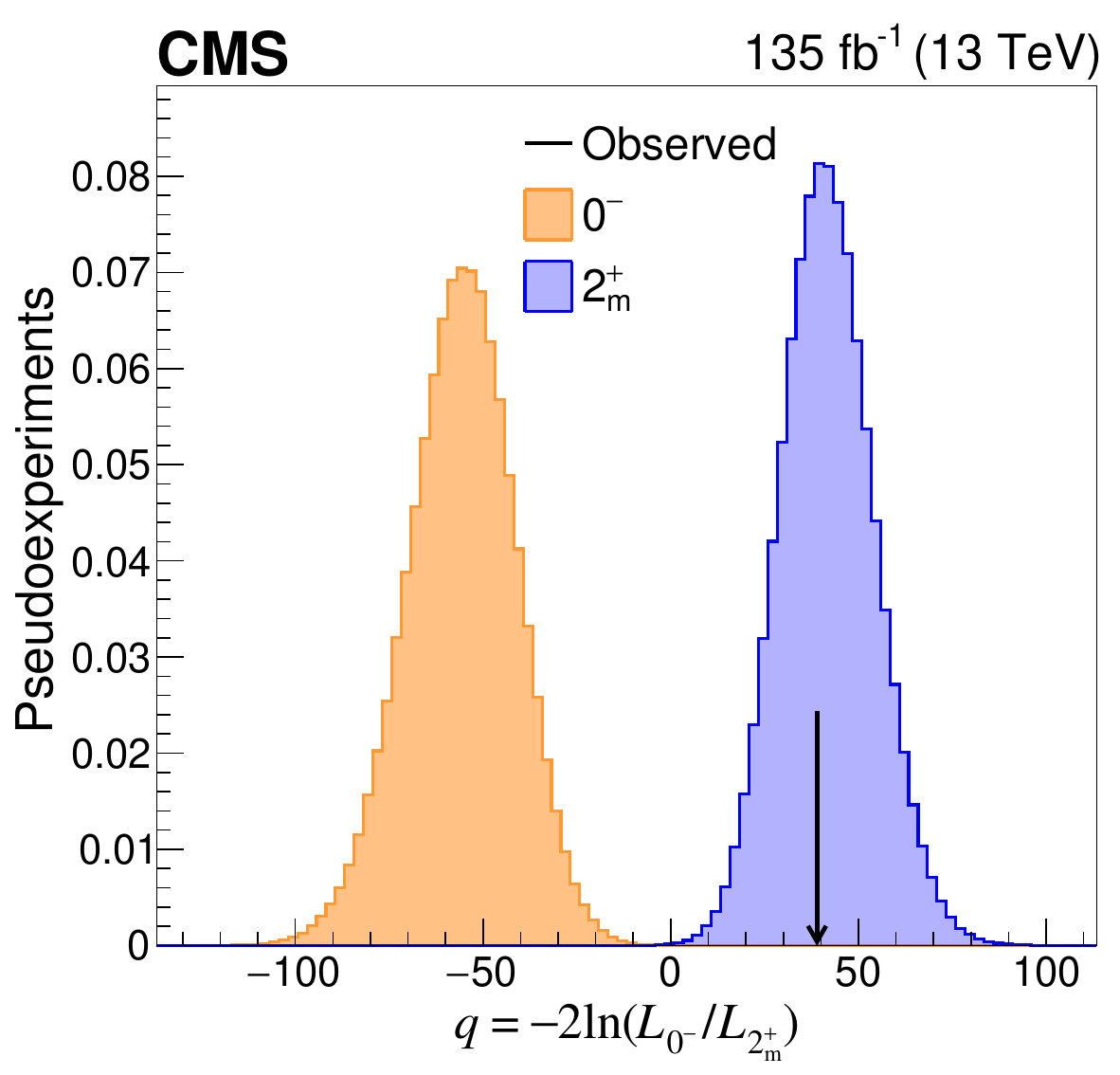}
    }
    \topcaption{
    \textbf{Analysis of angular distributions.}
    Left: Distributions of $\mathcal{D}_{2_m^{+}0^{-}}$ for the $0^{-}$, $2_m^{-}$, and $2_m^{+}$ models 
    in the range $6.2<m_{4\PGm}<8.0\GeV$.
    Distributions for signal only (dashed) and for signal plus background (solid and dash-dot-dotted) 
    models are compared to the experimental data points with error bars, with uncertainty bands 
    representing post-fit model uncertainties, which are partially correlated with the data.
    The $0^{-}$ and $2_m^{-}$ distributions are identical apart from systematic uncertainties arising from polarization effects.
    Right: Normalized distributions of the test statistic $q=-2{\ln(\mathcal{L}_{0^{-}}/\mathcal{L}_{2^{+}_{m}})}$ 
    from pseudoexperiments generated under the $2_m^{+}$ (blue/right) and $0^{-}$ (orange/left) hypotheses, 
    with the arrow indicating the observed value $q_\text{obs}$.
     }
    \label{fig:Dij}
\end{figure*}

We conduct pairwise comparisons between different models characterized by distinct $J^{PC}$ quantum numbers. 
Instead of analysing the multidimensional space of angular observables $(\theta_1, \theta_2, \Phi)$ directly, 
we define a single observable, $\mathcal{D}_{ij}$, constructed from the likelihood ratio between two models labelled $i$ and $j$.
It relies on quantum-mechanical calculations, as detailed in Appendix~\ref{sec:nature_append_mela} and Ref.~\cite{Bolognesi:2012mm}. 
This discriminant $\mathcal{D}_{ij}$ depends on the angular observables and is optimal for discriminating between the two hypotheses $i$ and $j$. 
The final statistical analysis of the data is carried out using two-dimensional distributions of events, $\mathcal{P}_{ij}(m_{4\PGm}, \mathcal{D}_{ij})$.
An example of the predicted distributions for the $\mathcal{D}_{2_m^{+}0^{-}}$ observable is presented in Fig.~\ref{fig:Dij} (left), 
where the $0^{-}$, $2_m^{-}$, and $2_m^{+}$ models are shown, along with the background and experimental data. 
Projecting the two-dimensional distribution $(m_{4\PGm}, \mathcal{D}_{ij})$ onto the single observable $\mathcal{D}_{ij}$
results in a partial loss of statistical power.

\section{Quantum number determination}
\label{sec:nature_results}

To distinguish alternative models, the test statistic $q=-2{\ln(\mathcal{L}_{J_j^P}/\mathcal{L}_{J_i^P})}$ 
is defined using the likelihood ratio of signal plus background likelihoods for the two signal hypotheses $J^P_i$ and $J^P_j$, 
based on the two observables $(m_{4\PGm}, \mathcal{D}_{ij})$, as outlined in Appendix~\ref{sec:nature_append_stat}.
In a maximum likelihood fit, the likelihood function $\mathcal{L}$ is maximized with respect to the nuisance parameters, 
which include the yields of signal and background processes, as well as constrained parameters that account for 
systematic uncertainties. These uncertainties encompass variations in mass shapes and in the discriminants
used to model both signal and background components, as detailed in Appendix~\ref{sec:nature_append_syst}. 

An example of the observed value of $q$, along with the expected distributions for the $2_m^{+}$ and $0^{-}$ models, 
is shown in Fig.~\ref{fig:Dij} (right). 
The expectations for each model are derived from a large ensemble of pseudoexperiments, 
with systematic uncertainties from the parameterization included in the fit procedure.
These pseudoexperiments are generated according to the observed data shown in Fig.~\ref{fig:mass}, 
combined with the kinematic distributions specific to each model.
Given the observed value of $q$, the $2^{+}_m$ model is clearly preferred over $0^{-}$.
All the steps outlined above for testing the spin-parity properties of the \PX states closely follow the methods 
and tools developed for the discovery and characterization of the Higgs boson in its four-lepton decay 
in 2012~\cite{Bolognesi:2012mm,CMS:2012qbp,CMS:2012vby}.

\begin{figure*}[th!]
    \centering
    \includegraphics[width=0.99\linewidth]{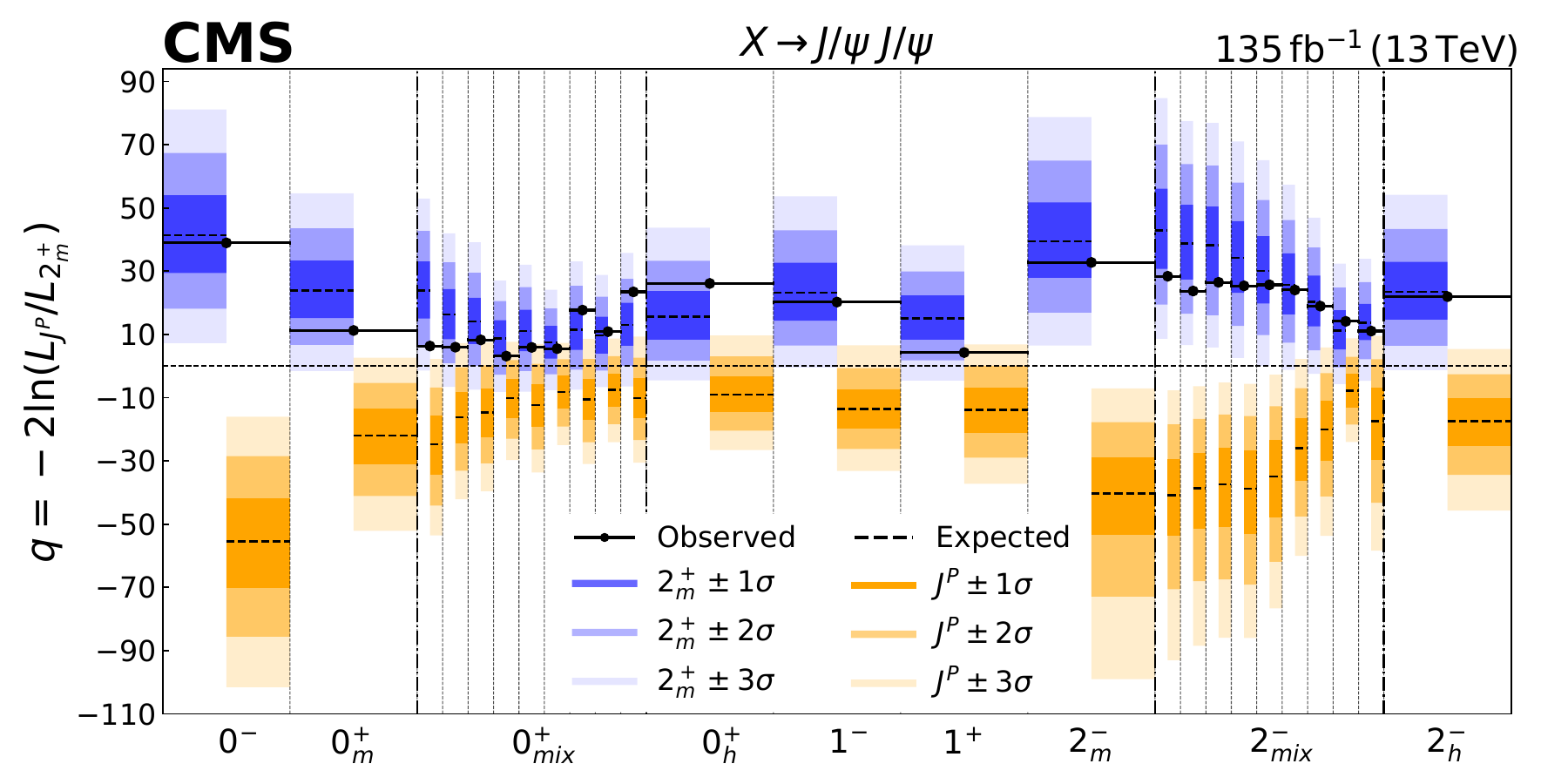}
    \topcaption{
     \textbf{Summary of statistical tests.}   
    Distributions of the test statistic $q$ for various $J_i^{P}$ hypotheses tested against the $2_m^{+}$ model. 
    The observed $q_\text{obs}$ values are indicated by the black dots.
    The expected median and the 68.3\%, 95.4\%, and 99.7\% confidence level regions for the $2_m^{+}$ model (blue/left) and 
    for each of the alternative $J_i^{P}$ hypotheses (orange/right) are shown. 
    The first entry corresponding to $0^-$ reflects the information shown in Fig.~\ref{fig:Dij} (right). 
    For $0^+$ and $2^-$ models, eleven points correspond to varying fractions in the mixture 
    of the two structures of interaction.
        }
     \label{fig:summary}
\end{figure*}

Following this methodology, all pairs of the eight $J_i^{P}$ models listed in Table~\ref{table:jcp} are tested,
and in each case involving the $2_m^{+}$ model, the latter is favoured.
The corresponding results are summarized in Fig.~\ref{fig:summary} and Table~\ref{table:results-short},
where the probability, $p$-value, and the associated $Z$-score, expressed as the number of standard 
deviations derived from the one-sided Gaussian tail integral, are shown for an alternative model $J_i^{P}$
tested against the $2_m^{+}$ model. 

Building on this observation, a mixture of the $0_m^{+}$ and $0_h^{+}$ structures of interactions, 
as well as that of $2_m^{-}$ and $2_h^{-}$, is tested in ten equal fractional increments against 
the $2_m^{+}$ model. The $2_m^{+}$ model is again preferred.
In each case, one of the mixed models shown in Fig.~\ref{fig:summary},
denoted as $0_{\text{mix}}^+$ or $2_{\text{mix}}^-$, represents the mixed scenario with the least 
separation from $2_m^{+}$ and is listed in Table~\ref{table:results-short}.
A mixture of $2_m^{-}$ and $2_h^{-}$ contributions does not produce interference because of their differing 
spin projections along the decay axis, as reflected in the amplitude composition presented in Table~\ref{table:jcp}.
We examine both constructive and destructive interference between the $0_m^{+}$ and $0_h^{+}$ 
models, by considering both positive and negative relative signs between their contributing amplitudes.
Constructive interference results in the smallest deviation from the $2_m^{+}$ model, except in the first 
$0_\text{mix}^{+}$ step shown in Fig.~\ref{fig:summary}, where the sign-induced model differences 
are minimal and destructive interference yields a slightly smaller separation. 
 
Based on the results presented in Fig.~\ref{fig:summary} and Table~\ref{table:results-short}, 
the tests for the $J^{PC}=0^{-+}$ and $1^{-+}$ scenarios reject these hypotheses
with a significance level exceeding $5$ standard deviations when compared to a $2_m^{+}$ model.
The $2^{-+}$ scenario, along with higher spin values that have the same $P$ and $C$ quantum numbers,
is excluded at a significance level of $3$ standard deviations. This establishes the quantum numbers  
$P=+1$ and $C=+1$, as shown by the decay final-state particles and their distributions.

\begin{table}[t!h]
    \centering
    \topcaption{
         \textbf{Summary of statistical tests.}   
         The $p$-value and the associated $Z$-score are shown for alternative models $J_i^{P}$, tested against the $2_m^{+}$ model. 
         A higher Z-score implies that the model is less compatible with observation.
     }
    \begin{tabular}{llc}
    \hline
  $J_i^{P}$   & ~~\,$p$-value & $Z$-score \\[-4pt]
     &   & reject $J_i^{P}$ \\
    \hline
$0^{-}$ & $2.7 \times 10^{-13}$ & $7.2$ \\[\cmsTabSkip]
$0_{m}^{+}$ & $4.3 \times 10^{-5}$ & $3.9$ \\
$0_{\text{mix}}^{+}$ & $1.4 \times 10^{-2}$ & $2.2$ \\
$0_{h}^{+}$ & $3.1 \times 10^{-9}$ & $5.8$ \\[\cmsTabSkip]
$1^{-}$ & $8.0 \times 10^{-8}$ & $5.2$ \\[\cmsTabSkip]
$1^{+}$ & $4.7 \times 10^{-3}$ & $2.6$ \\[\cmsTabSkip]
$2_{m}^{-}$ & $4.1 \times 10^{-12}$ & $6.8$ \\
$2_{\text{mix}}^{-}$ & $6.5 \times 10^{-4}$ & $3.2$ \\
$2_{h}^{-}$ & $2.2 \times 10^{-8}$ & $5.5$ \\
\hline
    \end{tabular}
    \label{table:results-short}
\end{table}

The $1^{++}$ scenario is excluded at more than 99\% confidence level, when compared to the $2_m^{+}$ model. 
The $J^{PC}=0^{++}$ scenario, when considering a combination of possible amplitudes, 
is excluded at more than 95\% confidence level, when compared to the $2_m^{+}$ model.
It is important to emphasize that the selected $2_m^{+}$ model represents just one possible realization 
of the $J^{PC}=2^{++}$ scenario, and an admixture of other amplitudes could lead to 
angular distributions resembling those of the $0^{++}$ or $1^{++}$ scenarios. 
The $J\ge3$ quantum numbers are still possible, but $J=2$ is more likely, due to the additional energy 
needed to achieve a higher angular excitation of the hadronic states with $L\ge2$.
This makes the $J^{PC}=2^{++}$ interpretation preferred for the fully charmed tetraquark states 
$\PXP{6600}$, $\PXP{6900}$, and $\PXP{7100}$.

\section{Discussion}
\label{sec:nature_discussion}

The study of tetraquark states has attracted considerable interest due to its potential to provide insight
into the structure of the hadronic matter that makes up the world around 
us~\cite{Berezhnoy:2011xn,
Wu:2016vtq,Karliner:2016zzc,
Richard:2017vry,Anwar:2017toa,Anwar:2017toa,
Nielsen:2018ytt,Nielsen:2018uyn,Ali:2019roi,Brambilla:2019esw,Liu:2019zuc,Wang:2019rdo,Bedolla:2019zwg,
liu:2020eha,Jin:2020jfc,Becchi:2020uvq,Chen:2020xwe,Wang:2020wrp,Dong:2020nwy,
Zhang:2020hoh,Zhu:2020xni,Liu:2021rtn,Faustov:2022mvs,
Feng:2023agq,Celiberto:2024mab,Belov:2024qyi,Zhu:2024swp}.
In this paper, we have presented the first measurement of the quantum numbers for a recently discovered 
family of three all-charm tetraquarks, based on data collected by the CMS experiment at the LHC.
Our results, summarized in Table~\ref{table:results-short}, favour a $J^{PC}=2^{++}$ assignment.

For a system of two constituents, as represented by either model in Fig.~\ref{fig:decay},  
an orbital angular momentum of $L = 1$ is excluded by the requirement $P = +1$, 
while higher orbital excitations with $L \geq 2$ are energetically disfavoured. 
This makes the $S$-wave ($L = 0$) configuration the most likely. 
In a molecular scenario, the quark-antiquark pairs are not required to be spin-1 mesons,
making a $J = 2$ configuration less likely. This picture agreed with existing data for all other well-established 
tetraquark candidates with measured spin, such as the $\PXP{3872}$ and $\PZ_{\PQc}(3900)^{+}$,
all of which have $J<2$~\cite{ParticleDataGroup:2024cfk}.
In contrast, a tightly-bound $\PQc\PQc\PAQc\PAQc$ tetraquark with a diquark--antidiquark configuration requires 
both diquarks to be in spin-1 states, which restricts the total spin to $J = 0$ or $J = 2$, making $J = 2$ a natural choice. 
This spin-1 diquark requirement, however, does not apply to tetraquark candidates with mixed-flavour quark content,
a consideration relevant to all previously observed candidates, which contained both heavy and light quarks~\cite{ParticleDataGroup:2024cfk}.
Other theoretical considerations also favour $J = 2$ for the tightly-bound tetraquark states~\cite{Becchi:2020uvq,Belov:2024qyi}.

This advancement in understanding exotic hadrons was enabled by the study of all-heavy tetraquarks
and it brings us closer to uncovering their internal structure.
Although our findings do not definitively distinguish between tightly-bound tetraquark and 
meson-meson molecular models, they provide constraints on the possible internal structures  
and favour the tightly-bound scenario.

\begin{acknowledgments}

We congratulate our colleagues in the CERN accelerator departments for the excellent performance of the LHC and thank the technical and administrative staffs at CERN and at other CMS institutes for their contributions to the success of the CMS effort. In addition, we gratefully acknowledge the computing centres and personnel of the Worldwide LHC Computing Grid and other centres for delivering so effectively the computing infrastructure essential to our analyses. We also acknowledge the Advanced Research Computing at Hopkins (ARCH) for providing computing resources essential for this analysis. Finally, we acknowledge the enduring support for the construction and operation of the LHC, the CMS detector, and the supporting computing infrastructure provided by the following funding agencies: SC (Armenia), BMBWF and FWF (Austria); FNRS and FWO (Belgium); CNPq, CAPES, FAPERJ, FAPERGS, and FAPESP (Brazil); MES and BNSF (Bulgaria); CERN; CAS, MoST, and NSFC (China); MINCIENCIAS (Colombia); MSES and CSF (Croatia); RIF (Cyprus); SENESCYT (Ecuador); ERC PRG, RVTT3 and MoER TK202 (Estonia); Academy of Finland, MEC, and HIP (Finland); CEA and CNRS/IN2P3 (France); SRNSF (Georgia); BMBF, DFG, and HGF (Germany); GSRI (Greece); NKFIH (Hungary); DAE and DST (India); IPM (Iran); SFI (Ireland); INFN (Italy); MSIP and NRF (Republic of Korea); MES (Latvia); LMTLT (Lithuania); MOE and UM (Malaysia); BUAP, CINVESTAV, CONACYT, LNS, SEP, and UASLP-FAI (Mexico); MOS (Montenegro); MBIE (New Zealand); PAEC (Pakistan); MES and NSC (Poland); FCT (Portugal); MESTD (Serbia); MICIU/AEI and PCTI (Spain); MOSTR (Sri Lanka); Swiss Funding Agencies (Switzerland); MST (Taipei); MHESI and NSTDA (Thailand); TUBITAK and TENMAK (Turkey); NASU (Ukraine); STFC (United Kingdom); DOE and NSF (USA).

\hyphenation{Rachada-pisek} Individuals have received support from the Marie-Curie programme and the European Research Council and Horizon 2020 Grant, contract Nos.\ 675440, 724704, 752730, 758316, 765710, 824093, 101115353, 101002207, 101001205, and COST Action CA16108 (European Union); the Leventis Foundation; the Alfred P.\ Sloan Foundation; the Alexander von Humboldt Foundation; the Science Committee, project no. 22rl-037 (Armenia); the Fonds pour la Formation \`a la Recherche dans l'Industrie et dans l'Agriculture (FRIA-Belgium); the Beijing Municipal Science \& Technology Commission, No. Z191100007219010, the Fundamental Research Funds for the Central Universities, the Ministry of Science and Technology of China under Grant No. 2023YFA1605804, and the Natural Science Foundation of China under Grant No. 12061141002 (China); the Ministry of Education, Youth and Sports (MEYS) of the Czech Republic; the Shota Rustaveli National Science Foundation, grant FR-22-985 (Georgia); the Deutsche Forschungsgemeinschaft (DFG), among others, under Germany's Excellence Strategy -- EXC 2121 ``Quantum Universe" -- 390833306, and under project number 400140256 - GRK2497; the Hellenic Foundation for Research and Innovation (HFRI), Project Number 2288 (Greece); the Hungarian Academy of Sciences, the New National Excellence Program - \'UNKP, the NKFIH research grants K 131991, K 133046, K 138136, K 143460, K 143477, K 146913, K 146914, K 147048, 2020-2.2.1-ED-2021-00181, TKP2021-NKTA-64, and 2021-4.1.2-NEMZ\_KI-2024-00036 (Hungary); the Council of Science and Industrial Research, India; ICSC -- National Research Centre for High Performance Computing, Big Data and Quantum Computing and FAIR -- Future Artificial Intelligence Research, funded by the NextGenerationEU program (Italy); the Latvian Council of Science; the Ministry of Education and Science, project no. 2022/WK/14, and the National Science Center, contracts Opus 2021/41/B/ST2/01369 and 2021/43/B/ST2/01552 (Poland); the Funda\c{c}\~ao para a Ci\^encia e a Tecnologia, grant CEECIND/01334/2018 (Portugal); the National Priorities Research Program by Qatar National Research Fund; MICIU/AEI/10.13039/501100011033, ERDF/EU, "European Union NextGenerationEU/PRTR", and Programa Severo Ochoa del Principado de Asturias (Spain); the Chulalongkorn Academic into Its 2nd Century Project Advancement Project, and the National Science, Research and Innovation Fund via the Program Management Unit for Human Resources \& Institutional Development, Research and Innovation, grant B39G680009 (Thailand); the Kavli Foundation; the Nvidia Corporation; the SuperMicro Corporation; the Welch Foundation, contract C-1845; and the Weston Havens Foundation (USA).
   
\end{acknowledgments}

\bibliography{auto_generated}

\appendix

\section{Methods} 
\label{methods}

\subsection{Modelling of hadron properties}
\label{sec:nature_append_model}

To investigate the spin-parity properties of the resonant structure within the $\PJGy\,\PJGy$ invariant mass 
spectrum in the range of 6.2--8.0\GeV, we use an approach designed to minimize 
model dependence. This approach relies on the observed $\PJGy\,\PJGy$ invariant mass spectrum
and the momentum of the $\PJGy\,\PJGy$ system in both transverse and longitudinal directions with respect to the beam,
while remaining independent of the system's polarization by relying solely on decay angular information.
For a spin-zero state, polarization is not relevant. For states with nonzero spin, we assume the state is produced 
unpolarized, but vary the polarization to evaluate potential small residual effects on the decay angular distributions 
due to detector acceptance.

The analysis employs a model that ensures consistency with the observations made by CMS, 
by utilizing simulation adjustments to accurately capture the observed transverse and longitudinal motion,
and parameters of the resonances and backgrounds extracted from Ref.~\cite{CMS:2023owd} and shown in Fig.~\ref{fig:mass}.
The background arises from nonresonant contributions, single-parton scattering (SPS) and double-parton 
scattering, plus an empirical term parameterizing the background near the threshold~\cite{CMS:2023owd}.
The background is parameterized using MC simulation, with adjustments applied to better match the observed 
data in both the signal and sideband regions.

We start by considering the spin-0 hypothesis for the $\PX$ states, which are produced without polarization. 
The helicity amplitudes $A_{\lambda_1\lambda_2}$ of the two $\PJGy$ mesons are listed in Table~\ref{table:jcp}. 
For the pseudoscalar state with $J^P = 0^-$, the amplitudes satisfy $A_{++} = -A_{--}$ and $A_{00} = 0$. 
In contrast, for the scalar state with $J^P = 0^+$, both $A_{++} = A_{--}$ and $A_{00}$ amplitudes contribute, 
with no specific prediction for the relative magnitude of $A_{00}$.
We adopt the general amplitude approach~\cite{Gao:2010qx,Bolognesi:2012mm,Gritsan:2020pib},
where the spin-0 state amplitude can be written as a sum of three Lorentz-invariant structures,
\ifthenelse{\boolean{cms@external}}{
\begin{equation}
    \begin{aligned}
        A(\PX_{J=0} \to V_1V_2) &= 
        a_{1}(q^2) m_{V}^2 \epsilon_1^* \epsilon_2^* \\
        &+ a_{2}(q^2) f_{\PGm \nu}^{*(1)}f^{*(2),\PGm \nu}\\
        &+ a_{3}(q^2) f^{*(1)}_{\PGm \nu} {\tilde f}^{*(2),\PGm  \nu},
    \label{eq:fullampl-spin0} 
    \end{aligned}
\end{equation}
}{
\begin{equation}
        A(\PX_{J=0} \to V_1V_2) = 
        a_{1}(q^2) m_{V}^2 \epsilon_1^* \epsilon_2^* 
        + a_{2}(q^2) f_{\PGm \nu}^{*(1)}f^{*(2),\PGm \nu}
        + a_{3}(q^2) f^{*(1)}_{\PGm \nu} {\tilde f}^{*(2),\PGm  \nu},
    \label{eq:fullampl-spin0} 
\end{equation}
}
and where the field strength tensor of a vector boson $V_i$ with momentum $q_i$ and polarization vector 
$\epsilon_i$ is defined as $f^{(i),{\PGm \nu}} = \epsilon_i^{\PGm}q_i^{\nu} - \epsilon_{i}^\nu q_i^{\PGm}$,
the conjugate field strength tensor is ${\tilde f}^{(i),\PGm \nu} = 1/2 \epsilon^{\PGm \nu \alpha \beta} f_{\alpha \beta}$,
and ${q}=q_1+q_2$. 

Since the $\PX$ decay proceeds via the strong interaction, which conserves parity, the first two terms 
in Eq.~(\ref{eq:fullampl-spin0}) can be interpreted as interactions involving scalar particles, corresponding 
to models $0_{m}^{+}$ and $0_{h}^{+}$ with couplings $a_1$ and $a_2$, respectively. 
The third term represents the interaction of a pseudoscalar particle, $0^{-}$, associated with the coupling $a_3$.
At $m_{4\PGm}=6.9\GeV$, $\abs{A_{00}}^2=52\%$ in model $0_m^{+}$
and $\abs{A_{00}}^2=19\%$ in model $0_h^{+}$.
In the following, we analyse the two models, $0_{m}^+$ and $0_{h}^{+}$, separately. 
Additionally, we consider their combination, denoted as $0_\text{mix}^+$, in which the relative magnitudes 
and signs of the couplings $a_1$ and $a_2$ are varied.
A positive relative sign results in constructive interference, while a negative relative sign leads to destructive interference.

\begin{figure*}[th]
  \centering
    \includegraphics[width=0.6\linewidth]{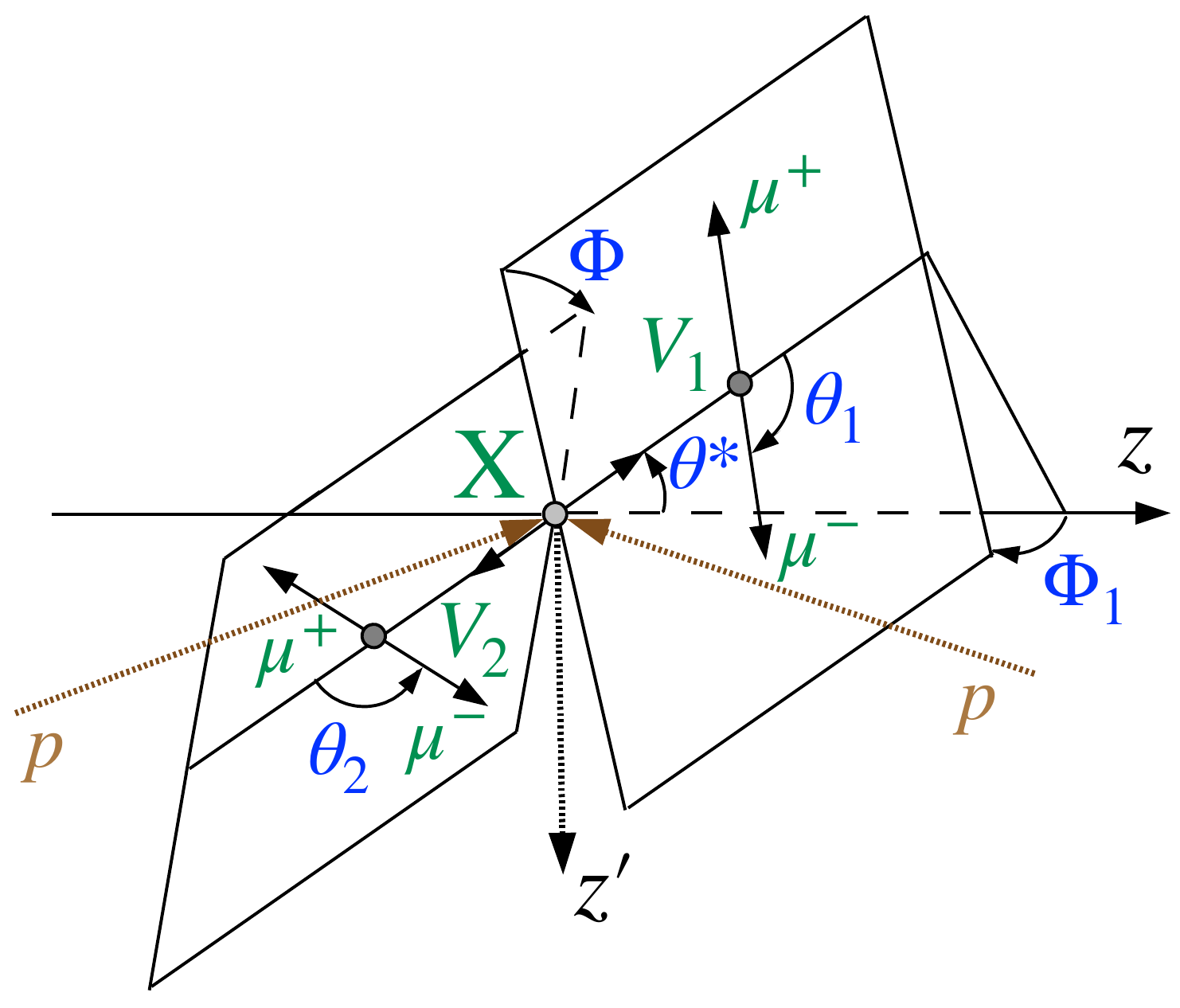}
    \topcaption{
\textbf{Angular observables.} 
The production and decay of a resonance $\PX$ in proton collisions $\Pp\Pp\to \PX \to \PV_1\PV_2 \to 4\PGm$ 
define the angular observables in the centre-of-mass frames of the corresponding particles~\cite{Gao:2010qx,Bolognesi:2012mm},
where the $\PV_1$ and $\PV_2$ refer to the $\PJGy$ mesons. The $z$ axis approximates the proton beam line,  
while the $z'$ axis corresponds to the direction of the four-muon system.
     }
    \label{fig:decay2}
\end{figure*}

The polarization of the spin-1 states varies depending on the production mechanism, 
where we consider either unpolarized production or polarization with $J_z$ or $J_{z'}=\pm 1$~\cite{Gao:2010qx}
because quark-initiated production dominates. The $z$ and $z'$ axes are defined in Fig.~\ref{fig:decay2},
where the motion of the four-muon system within the laboratory frame leads to appearance of noncollinear 
proton collisions and defines the $z'$ axis, while the $z$ axis approximates the proton beam line.
In the decay process, four helicity amplitudes contribute, and their relationships are shown in Table~\ref{table:jcp}.
This is equivalent to two Lorentz-invariant structures in the decay amplitude
\ifthenelse{\boolean{cms@external}}{
\begin{equation}
  \begin{aligned}
A(\PX_{J=1} \to V_1V_2) &= 
b_1(q^2) \big[  (\epsilon_1^* q) (\epsilon_2^* \epsilon_X) \\
  &+  (\epsilon_2^* q) (\epsilon_1^* \epsilon_X)  \big]\\
  &+  b_2(q^2) \epsilon_{\alpha\PGm\nu\beta} \epsilon_X^{\alpha} 
\epsilon_1^{*,\PGm} \epsilon_2^{*,\nu} \tilde{q}^\beta,
\label{eq:ampl-spin1} 
\end{aligned}
\end{equation}
}{
\begin{equation}
A(\PX_{J=1} \to V_1V_2) = 
b_1(q^2) \left[  (\epsilon_1^* q) (\epsilon_2^* \epsilon_X) 
  +  (\epsilon_2^* q) (\epsilon_1^* \epsilon_X)  \right]
+b_2(q^2) \epsilon_{\alpha\PGm\nu\beta} \epsilon_X^{\alpha} 
\epsilon_1^{*,\PGm} \epsilon_2^{*,\nu} \tilde{q}^\beta,
\label{eq:ampl-spin1} 
\end{equation}  
}
where $\tilde{q}=q_1-q_2$ and $\epsilon_\PX$ is the polarization vector of the spin-1 resonance $\PX$.
The $b_1$ and $b_2$ are the couplings to a vector state with $J^P = 1^-$ and an axial vector with $J^P = 1^+$, respectively.

The polarization of the spin-2 states also depends on the production mechanism, where we consider 
either unpolarized production or polarization with $J_z$ or $J_{z'}=0$ or $\pm 2$~\cite{Gao:2010qx}
because gluon-initiated production is expected to dominate. 
In the decay process, all nine helicity amplitudes $A_{\lambda_1\lambda_2}$ contribute.
This corresponds to the Lorentz-invariant structures in the decay amplitude found in Refs.~\cite{Gao:2010qx,Bolognesi:2012mm}.
There are two degrees of freedom in the $2^{-+}$ case, and we use the $2_m^-$ and $2_h^-$ models, 
corresponding to $g^{(2)}_{8}\ne 0$ and $g^{(2)}_{10}\ne 0$ in Refs.~\cite{Gao:2010qx,Bolognesi:2012mm}, respectively. 
The $2_m^-$ model corresponds to $A_{++}=-A_{--}$, with decay angular distributions identical to those of the $0^-$ case. 
For $2_h^-$, we have $A_{+0}=A_{0+}=-A_{-0}=-A_{0-}$, with decay angular distributions identical to those of the $1^-$ case. 

We employ a single representative model for the $J^{PC}=2^{++}$ state, corresponding to 
model $g^{(2)}_{1}=g^{(2)}_{5}\ne 0$ denoted $2_m^+$ in Refs.~\cite{Gao:2010qx,Bolognesi:2012mm}.
This $2_m^+$ model is chosen because the composition of $A_{\lambda_1\lambda_2}$ amplitudes represents 
all possible polarizations, in particular those which are unique for spin-2, $A_{+-}$ and $A_{-+}$, and the
tensor structure of interactions is minimal, avoiding the inclusion of higher-dimension operators.   
At $m_{4\PGm}=6.9\GeV$, $2\abs{A_{++}}^2=9\%$, $\abs{A_{00}}^2=21\%$, $4\abs{A_{+0}}^2=47\%$, 
and $2\abs{A_{+-}}^2=23\%$ in model $2_m^{+}$.
Therefore, if the data are consistent with $J^P=2_m^+$ and not with the other models, it will provide an unambiguous 
determination of $J\ge2$. A higher spin scenario ($J>2$) could also be possible, and it would exhibit angular 
distributions similar to those of $J=2$ with the same parity~\cite{Gao:2010qx,Bolognesi:2012mm}. 

Since the $J^{PC}=2^{++}$ state has four degrees of freedom in the amplitude composition, three other possibilities 
could be chosen that lead to the same observable decay distributions as the $0_m^+$, $0_h^+$, and $1^+$ states. 
Interference between the corresponding $2^{++}$ amplitude tensor structures is also possible. 
Therefore, from a purely decay-based analysis, if the data are consistent with either the $0^{++}$ or $1^{++}$ model, 
one cannot rule out a general  $2^{++}$  model. 
While analysing polarization information through production-sensitive angular distributions could help resolve this ambiguity, 
such an analysis is more challenging, due to uncertainty in the production mechanism.
If the resonance is produced unpolarized, such an analysis would not provide additional information.
In this paper, we check production-sensitive angular distributions for consistency with an unpolarized case, 
while a more detailed analysis is left for future work. 

Determining the form factors associated with the tensor structure of interactions, represented by 
$a_{i}(q^2)$ in Eq.~(\ref{eq:fullampl-spin0}) for spin-0, $b_{i}(q^2)$ in Eq.~(\ref{eq:ampl-spin1}) for spin-1, 
or equivalent $g^{(2)}_{i}(q^2)$ in Refs.~\cite{Gao:2010qx,Bolognesi:2012mm} for spin-2, 
with $q^2=m^2_{4\PGm}$, is nontrivial and relies on the model of strong interactions.
Nevertheless, it is completely separable from the computation of the angular distributions, 
given a particular tensor structure of interactions.
Furthermore, interference between the resonances could result in a complex alteration of the angular distributions. 
Nonetheless, assuming that all three resonances possess identical quantum numbers and coupling constants, 
the angular distributions remain unaffected by interference effects observed in the mass spectrum. 
Thus, in this study, we use an empirical approach to analyse the four-muon mass spectrum observed in the data, 
separating it from the analysis of angular distributions. This makes our analysis independent of the form factor model.
The form factors and interference effects are integrated into the observed mass spectrum, as shown in~Fig.~\ref{fig:mass}.

\subsection{Angular observables}
\label{sec:nature_append_angles}

The angular information used in the analysis is shown in Fig.~\ref{fig:decay2}~\cite{Gao:2010qx,Bolognesi:2012mm}.
Two production axes are defined, corresponding to the two production mechanisms,
either the direct short-distance production in parton collisions or the fragmentation of a single parton into the hadron. 
Axis $z$ is parallel to the beam line within the frame boosted along the beam line from the laboratory frame, 
where the longitudinal momentum of $\PX$ is zero, potentially reflecting polarization in the two-parton collisions. 
The $z'$ axis aligns with the direction of the $\PX$ momentum, 
reflecting potential polarization in the single-parton fragmentation scenario. 

The decay angles $\Phi$, $\theta_1$, and $\theta_2$ are defined with reference to Fig.~\ref{fig:decay}. 
The production angles $\theta^*$ and $\Phi_1$, or alternatively $\theta^{\prime*}$ and $\Phi_1'$, 
are defined with respect to the $z$ or $z'$ axis, respectively. 
The angle $\theta^*$ is defined between the $z$ axis and the $\PX$ decay axis in its rest frame,
while $\Phi_1$ is the angle between the first $\PJGy$ decay plane and the production plane, 
defined by the $z$ axis and the $\PX$ decay axis.
Figure~\ref{fig:decay2} illustrates only $\theta^*$ and $\Phi_1$, while $\theta^{\prime*}$ and $\Phi_1'$ are defined analogously.
These production angles are not used directly in the analysis to avoid dependence on the production model, 
but they are checked for consistency.

\begin{figure*}[ht!]
  \centering
    \centerline{
      \includegraphics[width=0.41\linewidth]{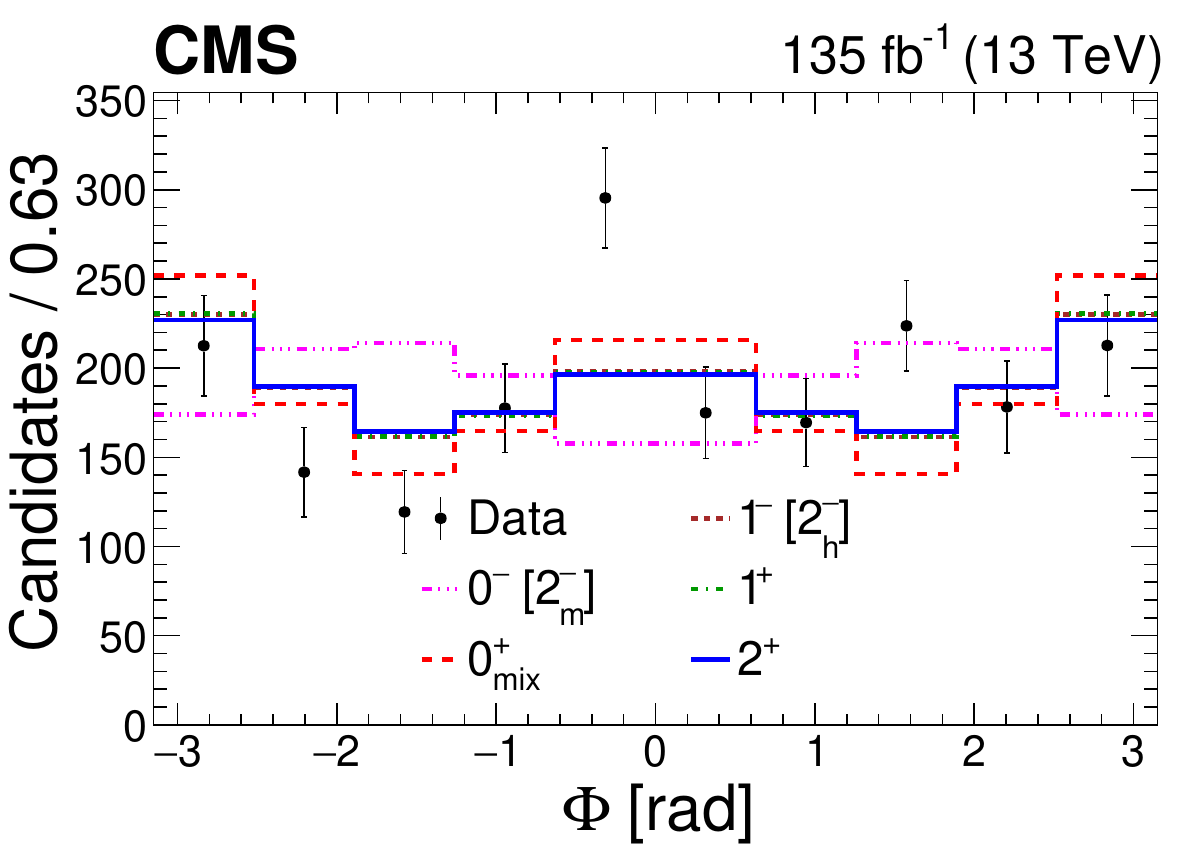}
    }
    \centerline{
      \includegraphics[width=0.41\linewidth]{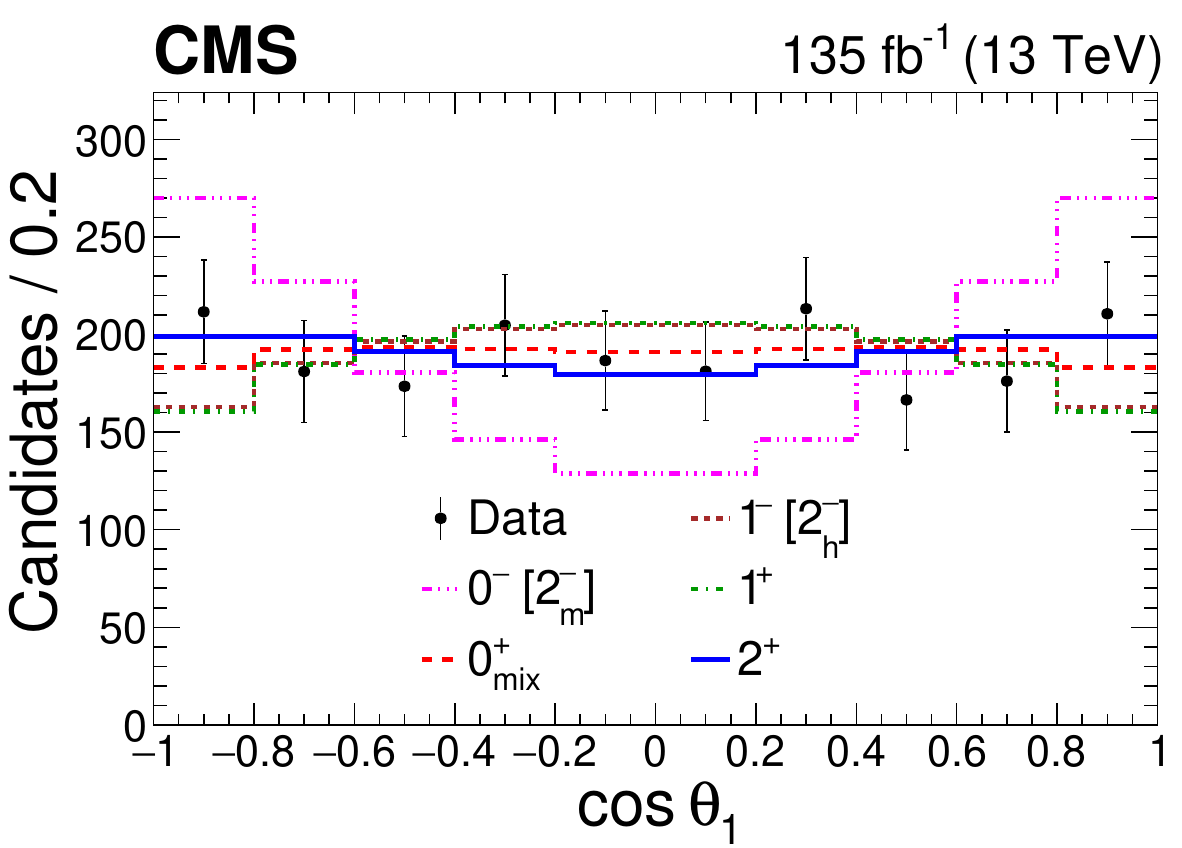}
      \includegraphics[width=0.41\linewidth]{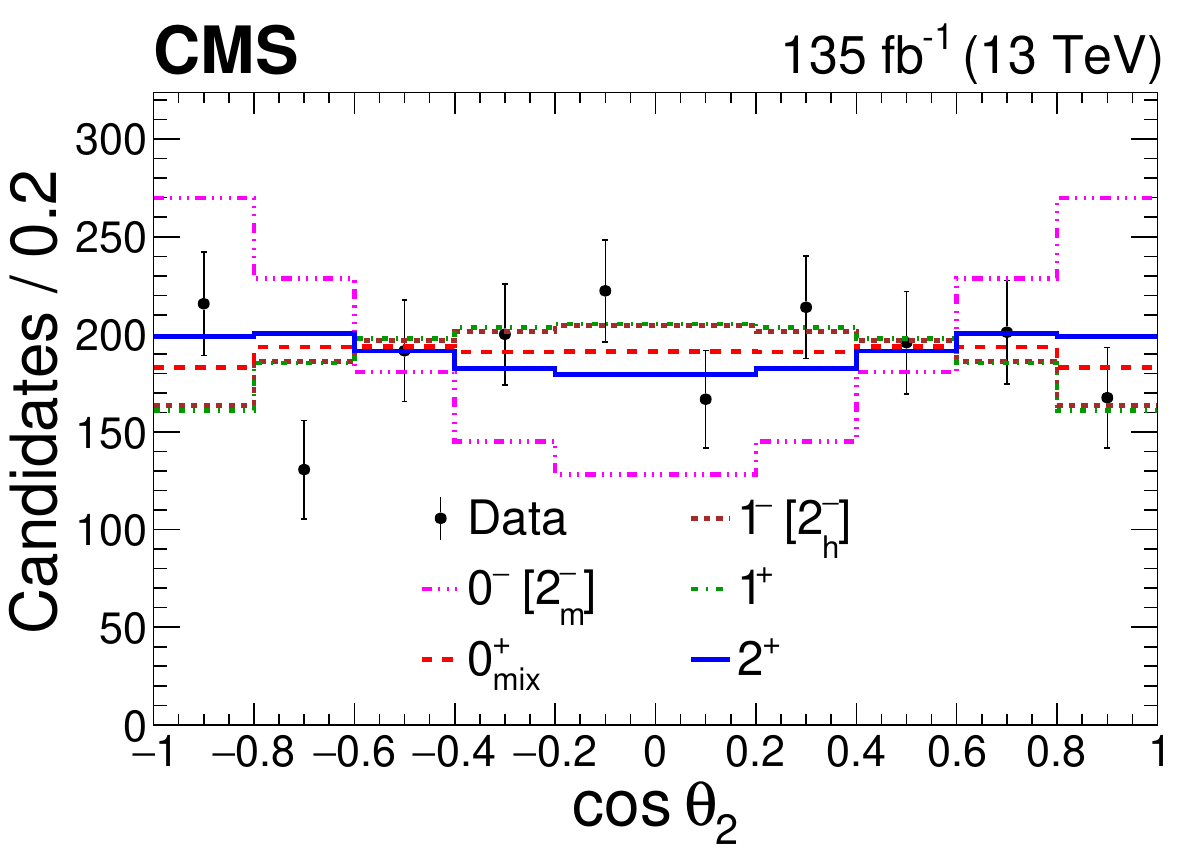}
    }
    \centerline{
      \includegraphics[width=0.41\linewidth]{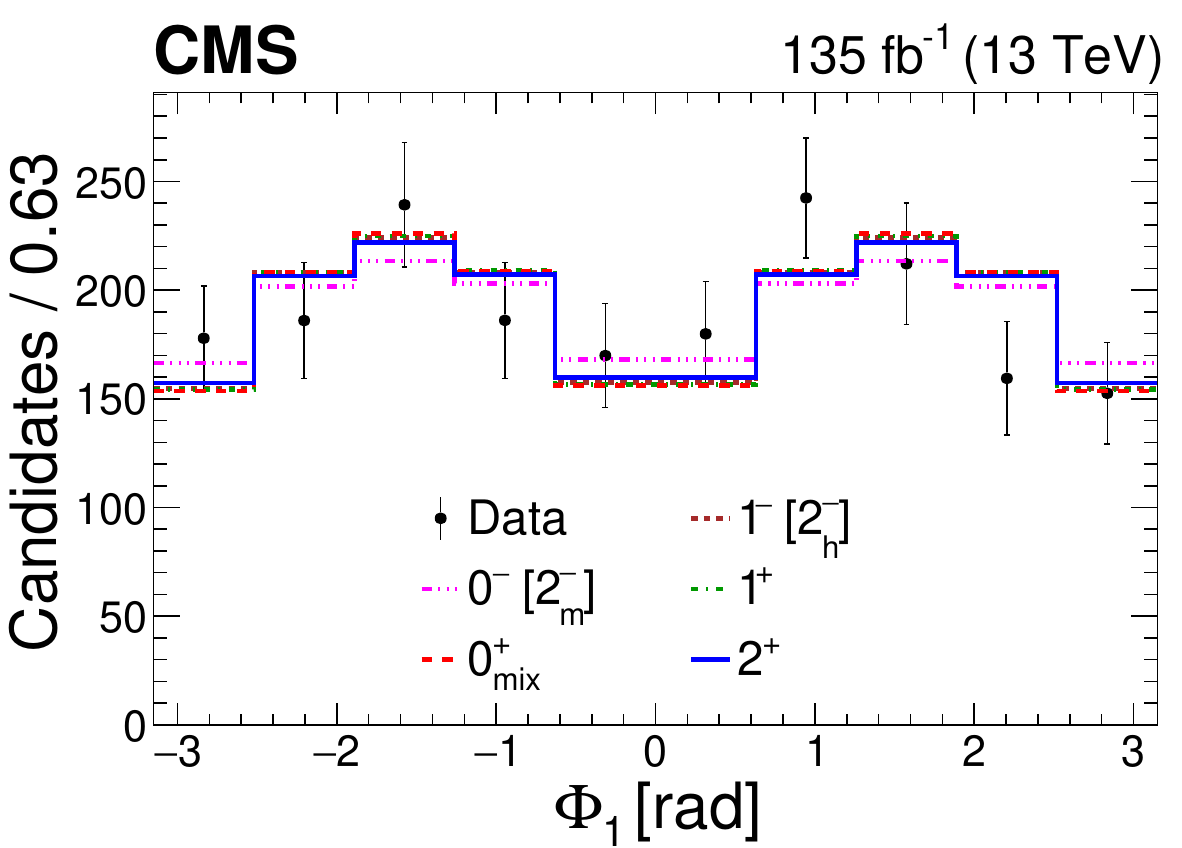}
      \includegraphics[width=0.41\linewidth]{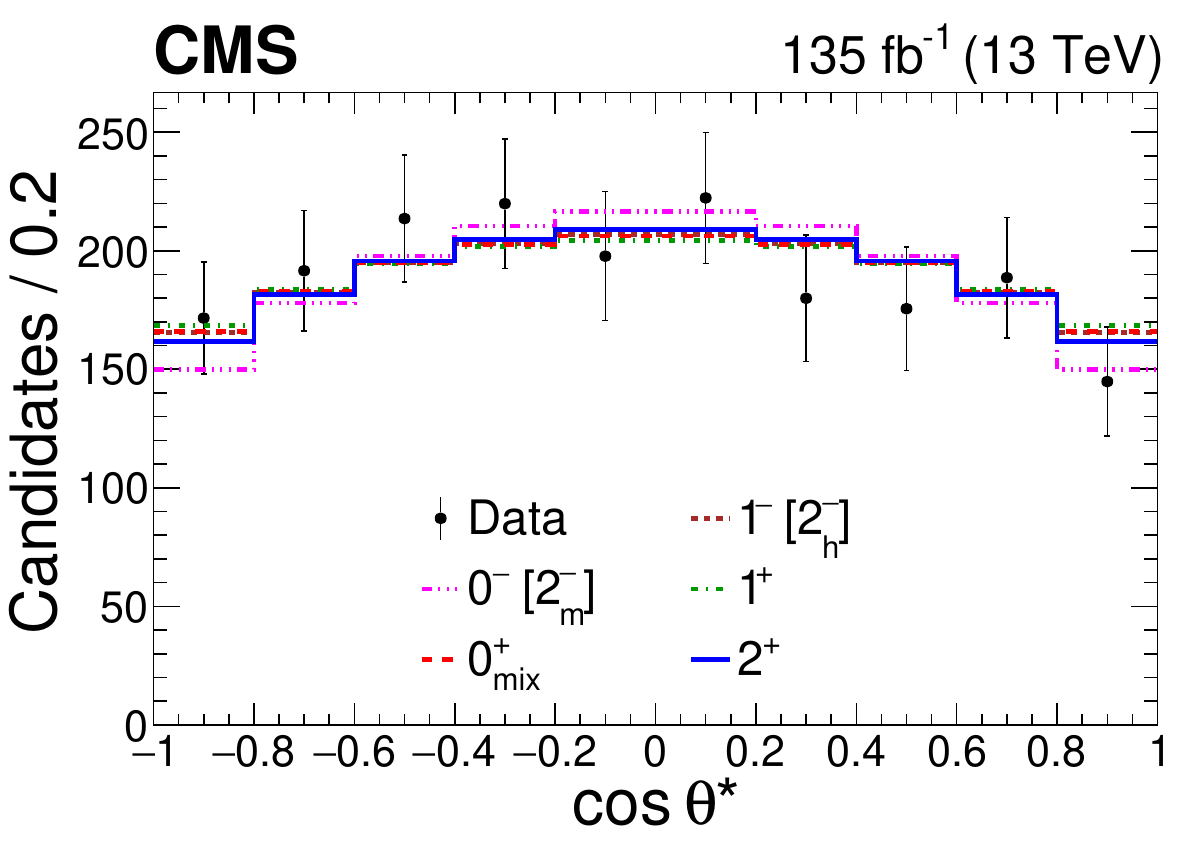}
    }
    \centerline{
      \includegraphics[width=0.41\linewidth]{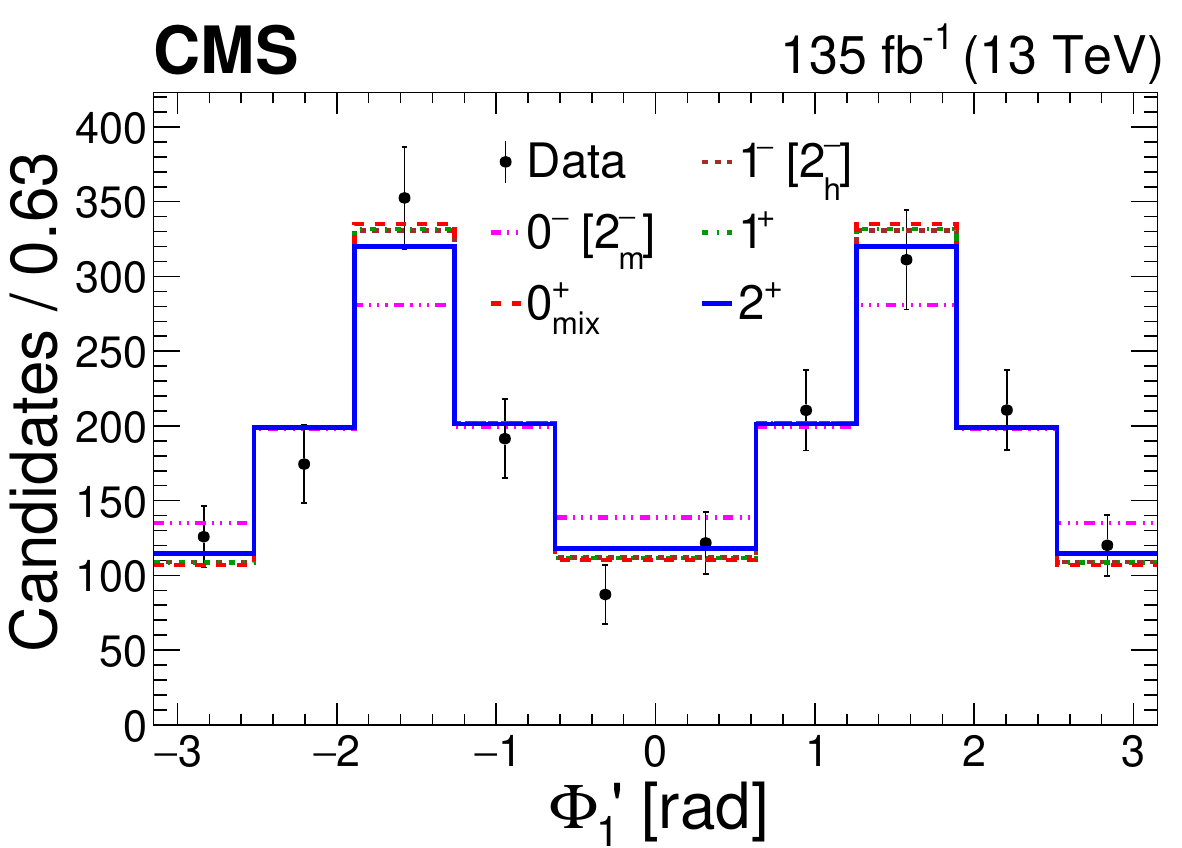}
      \includegraphics[width=0.41\linewidth]{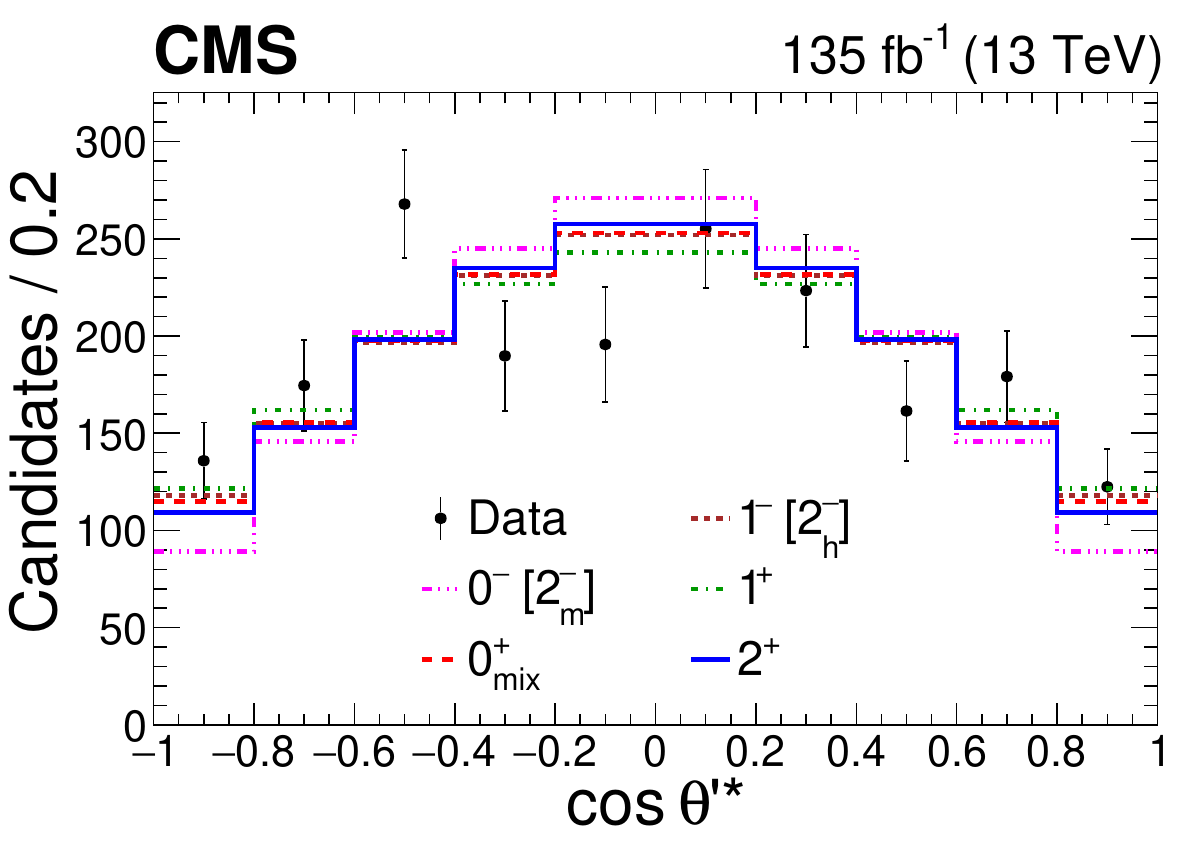}
    }
    \topcaption{
\textbf{Angular distributions.}
Distribution of the decay angles: $\Phi $ (upper row), $\cos\theta_1, \cos\theta_2$ (second row),
production angles defined with respect to axis $z$: $\Phi_1, \cos\theta^*$ (third row),
and defined with respect to axis $z'$: $ \Phi_1', \cos\theta^{\prime *}$ (lower row)
in the range $6.2<m_{4\PGm}< 8.0\GeV$ presented together with the five $J^P_i$ signal models.
The background is subtracted from the data, based on the expected distributions,
and systematic uncertainties are not incorporated in these plots.
The $0^{-}$ and $2_m^{-}$ distributions are identical, as are those of $1^{-}$ and $2_h^{-}$.
     }
    \label{fig:angles}
\end{figure*}

The distributions of the decay and production angles in the range $6.2<m_{4\PGm}< 8.0\GeV$ are presented
in Fig.~\ref{fig:angles}, together with the five signal models shown. 
It is important to note that the one-dimensional angular distributions in Fig.~\ref{fig:angles} 
do not capture all the information accessible to the optimal discriminant.
Correlations between angles are lost in the projections; for instance, the models $J^{PC}=1^{-+}$ and $1^{++}$ 
cannot be distinguished in each projection, but they can be separated using the optimal discriminants, 
which preserve all angular correlations.
Furthermore, Fig.~\ref{fig:angles} displays all events in the range 
$6.2<m_{4\PGm}< 8.0\GeV$ together, without accounting for the correlation between the angular distributions and $m_{4\PGm}$, 
or the variation in signal purity with $m_{4\PGm}$ due to the resonance structure appearing above the background.
As a result, the separation power shown in these illustration plots is considerably diminished compared to the full analysis.

To eliminate the dependence on the initial polarization of nonzero-spin states, 
the production angles are excluded from the data analysis.
The assumption is made that the resonances are produced unpolarized, 
although the variation of this polarization along the production axes is allowed for states with nonzero spin. 
This allows for the examination of any residual effects stemming from polarization dependence due to  
detector acceptance effects. 
Figure~\ref{fig:angles} shows that the production angular distributions are consistent with unpolarized resonance 
states along both the $z$ and $z'$ axes.

\subsection{Matrix element approach}
\label{sec:nature_append_mela}

The analysis of the multidimensional distributions $\mathcal{P}(\theta_1, \theta_2, \Phi, m_{4\PGm})$ 
is complicated by the complex description and the nontrivial effects of detector reconstruction.
Rather than utilizing the three angular observables directly, we construct a single observable
that effectively projects the angular information onto one dimension and is optimal for distinguishing 
between two hypotheses, $J^P_i$ and $J^P_j$. 
Utilizing a matrix element likelihood approach~\cite{Gao:2010qx,Bolognesi:2012mm,Gritsan:2020pib},
a kinematic discriminant is formulated based on the ratio of probabilities for hypotheses $J^P_i$ and $J^P_j$:
\begin{equation}
    \begin{aligned}
    \label{eq:melaSig}
{\mathcal{D}_{ij}}(m_{4\PGm},\vec\Omega)=\frac{\mathcal{P}_{j}(m_{4\PGm},\vec\Omega)}{\mathcal{P}_{i}(m_{4\PGm},\vec\Omega)+\mathcal{P}_{j}(m_{4\PGm},\vec\Omega)},
    \end{aligned}
\end{equation}
where $\mathcal{P}_{i}$ is the normalized probability based on the matrix element squared for a given hypothesis $J^P_i$. 
These matrix elements are computed within the same quantum-mechanical formalism as employed for the generation 
of MC events, as detailed in Section~\ref{sec:nature_analysis}. 

\begin{figure*}[t!]
  \centering
    \centerline{
      \includegraphics[width=0.48\linewidth]{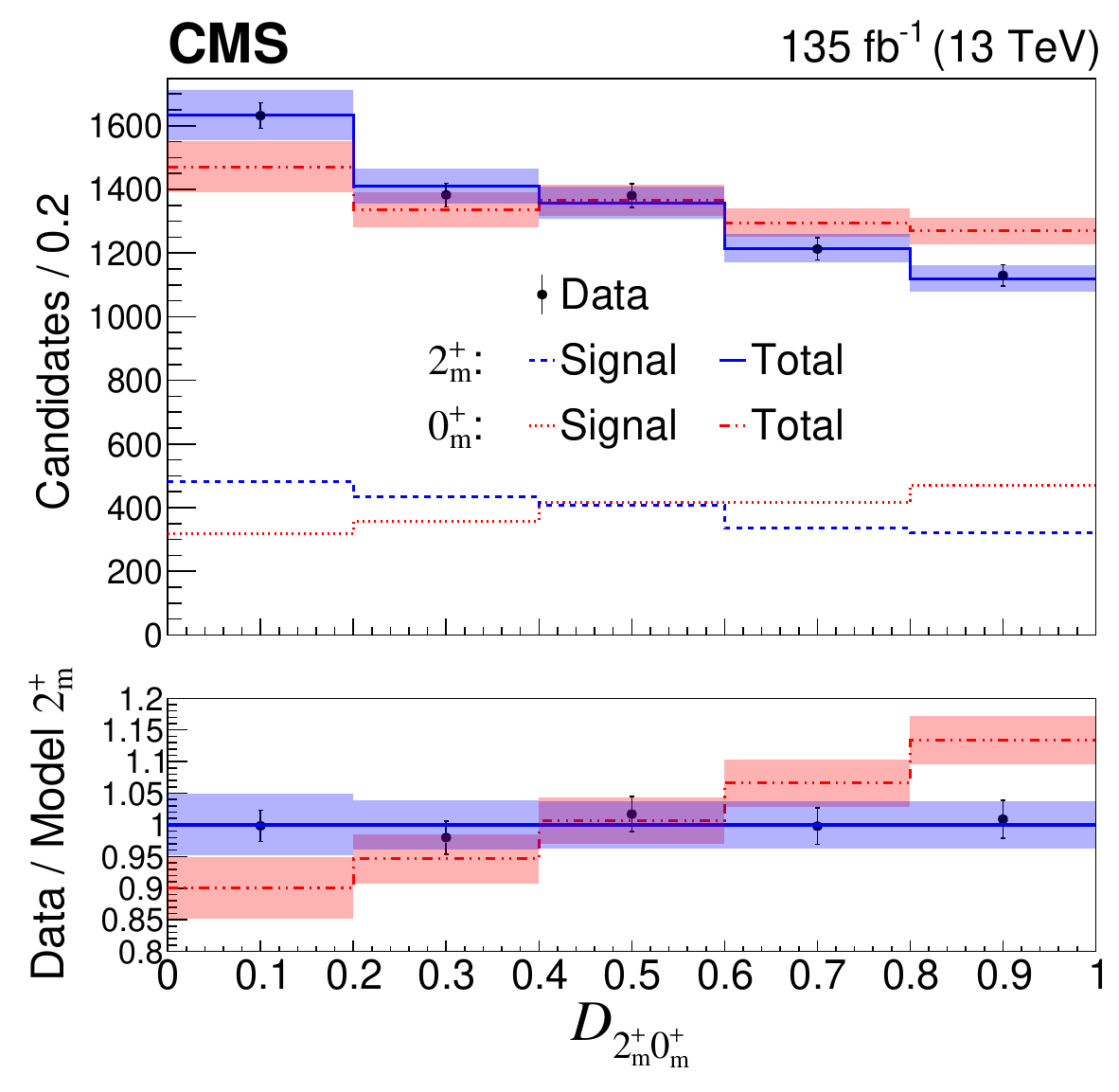}
       \includegraphics[width=0.48\linewidth]{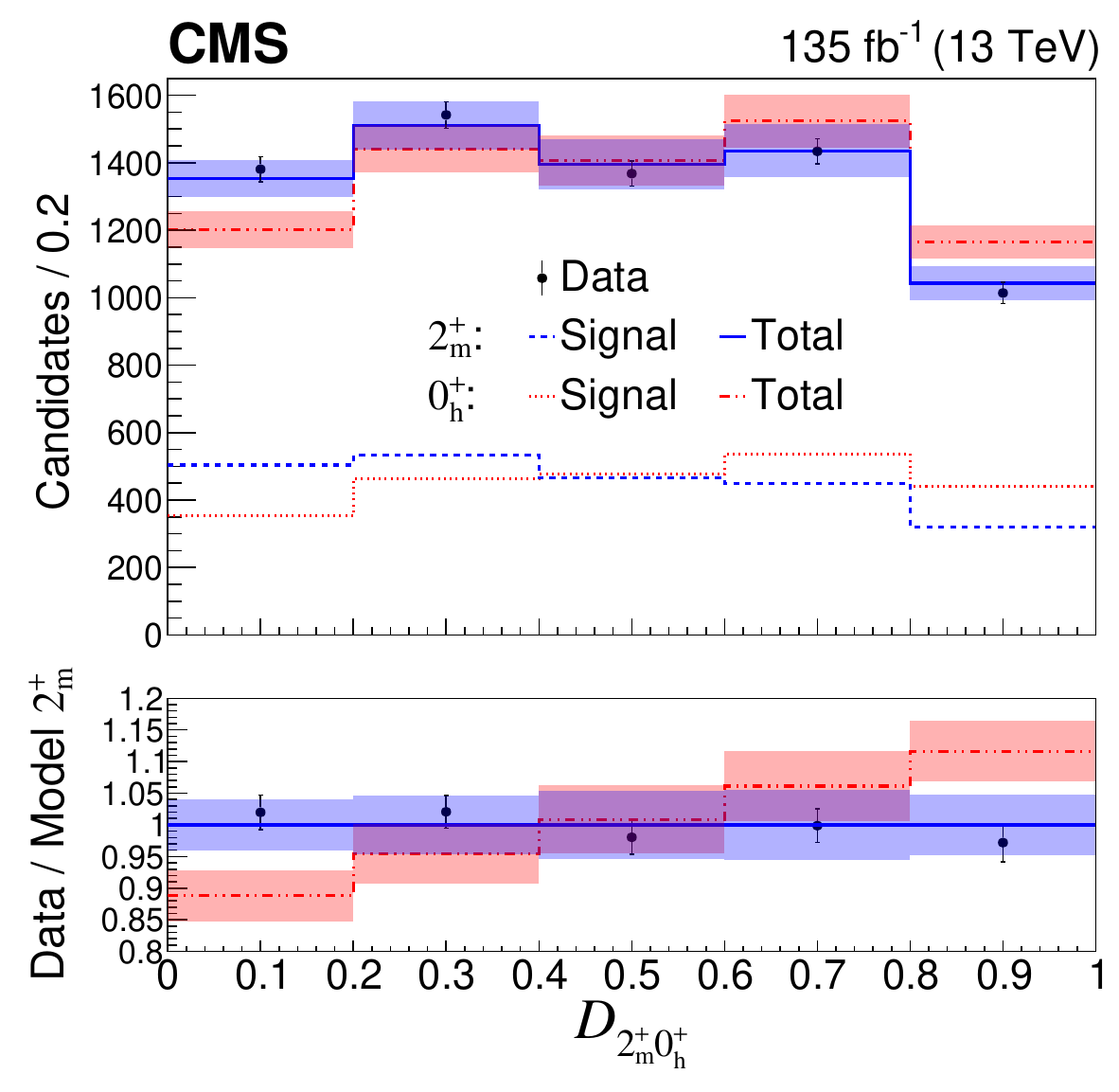}
    }
    \centerline{
      \includegraphics[width=0.48\linewidth]{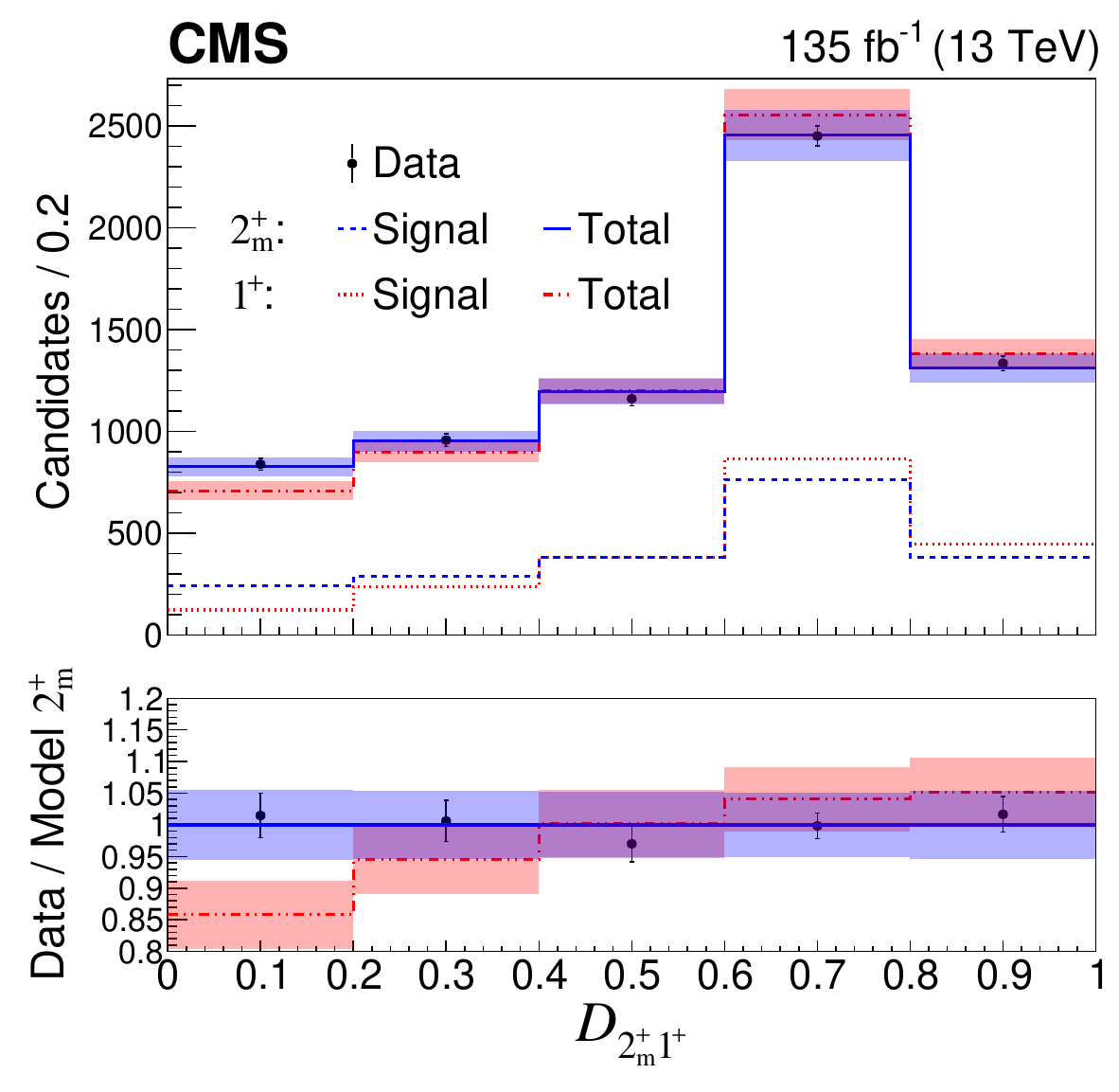}
      \includegraphics[width=0.48\linewidth]{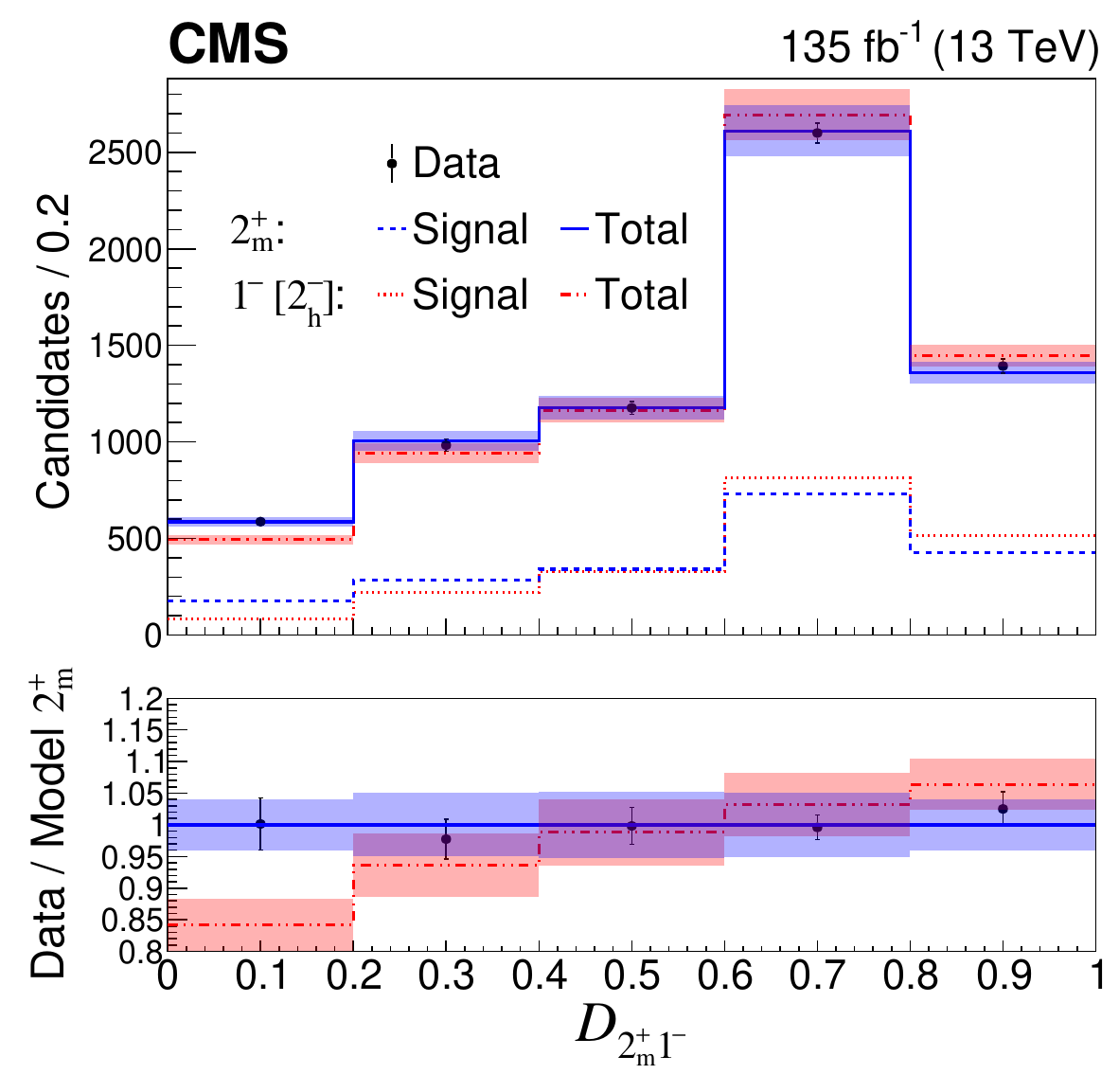}
    }
    \topcaption{
    \textbf{Optimal observables.}
Distributions of $\mathcal{D}_{ij}$ that are optimal for separating the $2_{m}^+$ model against the $0_{m}^+$ (upper left), 
$0_{h}^+$ (upper right), $1^-$ (lower left), and $1^+$ (lower right) models in the range $6.2<m_{4\PGm}<8.0\GeV$.
Distributions for signal only (dashed) and for signal plus background (solid and dash-dot-dotted) 
models are compared to the experimental data points with error bars, with uncertainty bands 
representing post-fit model uncertainties, which are partially correlated with the data.
The $1^{-}$ and $2_h^{-}$ distributions are identical, apart from systematic uncertainties arising from polarization effects.
The lower panels display the ratios of the data and of the model predictions to the mean expectations from the $2_{m}^+$ model.
     }
     \label{fig:discriminants}
\end{figure*}

In Eq.~(\ref{eq:melaSig}), we use only the decay angles $\vec\Omega=\{\cos\theta_1, \cos\theta_2, \Phi \}$.
The analysis is conducted using the two-dimensional distributions of $(m_{4\PGm}, \mathcal{D}_{ij})$.
Production information, including $\{\cos\theta^*, \Phi_1 \}$ or $\{\cos\theta^{\prime*}, \Phi^{\prime}_1 \}$, 
can be incorporated into a future analysis if a study of the resonance polarization is desired. 
The distributions of the discriminants, which were calculated to assess alternative models in comparison to the $2_m^+$ model, 
are presented in Figs.~\ref{fig:Dij} and~\ref{fig:discriminants} for all models defined without considering amplitude mixtures.
The kinematic distributions in Figs.~\ref{fig:Dij}, \ref{fig:angles}, and~\ref{fig:discriminants}  have also been separately examined 
in three different ranges of $m_{4\PGm}$, each corresponding to one of the three resonance structures depicted in Fig.~\ref{fig:mass}.
The data remain consistent with the $2_{m}^+$ model across the full mass range as well as within each of the three individual intervals.

The final statistical analysis of the data is carried out using two-dimensional distributions of events, $\mathcal{P}_{ij}(m_{4\PGm}, \mathcal{D}_{ij})$, 
where the predicted $m_{4\PGm}$ distribution is modelled in the same way as in Fig.~\ref{fig:mass} following Ref.~\cite{CMS:2023owd},
and the $\mathcal{D}_{ij}$ distribution is obtained in bins of 0.05\GeV in $m_{4\PGm}$ from the detailed MC simulation discussed 
earlier in Section~\ref{sec:nature_analysis}.
For any given slice of $m_{4\PGm}$, only five bins of $\mathcal{D}_{ij}$ are used, meaning that the MC simulation 
can accurately predict probability distributions using an affordable number of simulated events.

\subsection{Systematic uncertainties}
\label{sec:nature_append_syst}

In the maximum likelihood fit introduced in Sec.~\ref{sec:nature_results} and further detailed in 
Sec.~\ref{sec:nature_append_stat}, the likelihood function is maximized with respect to the nuisance 
parameters, which include those representing systematic uncertainties. 
The fit results encompass systematic variations in the parameterization of both signal and background models. 
The yields of the signal and each of the background processes are treated as free parameters, 
to be fully determined by the fit to the data.
Further variations are categorized into two groups: variations in mass shapes and variations in discriminants, 
applied to both signal and background components.

The study of systematic variations in mass shapes was detailed in Ref.~\cite{CMS:2023owd}, which includes 
resonance parameterization, resolution, and efficiency for signal components, as well as different models 
for background components. All these variations in mass shapes were incorporated into the two-dimensional 
analysis presented here, through the variation of $\mathcal{P}_k(m_{4\PGm})$ in Eq.~(\ref{eq:pdf_2D}).
Therefore, all uncertainties associated with the parameterization of the $m_{4\PGm}$ spectrum as detailed 
in Ref.~\cite{CMS:2023owd} are included in the analysis.

One of the uncertainties arises from the unknown angular distributions of the 
background contribution near the kinematic threshold~\cite{CMS:2023owd}. 
To allow maximum flexibility, the discriminant parameterization of this contribution, $T_{ijk}(\mathcal{D}_{ij}|m_{4\PGm})$ 
in Eq.~(\ref{eq:pdf_2D}), is expressed as the sum of the SPS background model and the two signal models under
investigation, with the relative contributions of all three components determined by freely floating nuisance parameters.
This approach is intended to accommodate a broad range of hypotheses for the threshold contribution, 
including possible variations of the nonresonant background and potential resonance excitations.

Other systematic variations in discriminants account for possible inconsistencies in \pt and $p_{z}$ distributions 
between data and MC simulation. The \pt and $p_{z}$ distributions are validated and tuned through 
comparisons between data and simulation in both the sideband region (8.0--9.0\GeV) and the 
signal region (6.2--8.0\GeV). Residual discrepancies in the discriminants due to mismodeling 
of \pt and $p_{z}$ are addressed through nuisance parameters, which allow for alternative 
discriminant parameterizations of $T_{ijk}(\mathcal{D}_{ij}|m_{4\PGm})$ in Eq.~(\ref{eq:pdf_2D}).
Similarly, uncertainties in the kinematic distributions of the nonresonant background components 
are evaluated through comparisons between data and MC simulations in the sideband region, 
where a satisfactory agreement is observed.  
Remaining discrepancies in the discriminants are handled through nuisance parameters 
that permit alternative discriminant parameterizations of $T_{ijk}(\mathcal{D}_{ij} \mid m_{4\PGm})$.

The nominal discriminant parameterization of all spin-1 and spin-2 resonances assumes no polarization. 
However, to evaluate the impact of detector acceptance effects on the results, we introduce an alternative 
parameterization that assumes polarization of spin-1 and spin-2 resonances along either the $z$ or $z'$ axes. 
This approach models polarized production arising from either two-parton collisions or single-parton fragmentation.

\subsection{Statistical analysis}
\label{sec:nature_append_stat}

We perform a binned extended maximum likelihood fit where the probability density function is a sum of contributions 
from all signal and background processes implemented in the Combine tool (v10.0.2)~\cite{CMS:2024onh}. This method mirrors the approach used to determine 
the spin-parity quantum numbers of the Higgs boson at the LHC~\cite{CMS:2014nkk}, as detailed in Ref.~\cite{Bolognesi:2012mm}.
Each process $k$ is characterized by a probability density function $\mathcal{P}_{ijk}$, used to analyse signal hypotheses $i$ and~$j$.
This function depends on two observables, $m_{4\PGm}$ and $\mathcal{D}_{ij}$, and is defined as a template binned in a $36{\times}5$ grid:
\begin{align}
  \mathcal{P}_{ijk}(m_{4\PGm}, \mathcal{D}_{ij}) = \mathcal{P}_k(m_{4\PGm}) T_{ijk}(\mathcal{D}_{ij}|m_{4\PGm}),
  \label{eq:pdf_2D}
\end{align}
where $ \mathcal{P}_k(m_{4\PGm})$ represents the probability density function of the invariant mass $m_{4\PGm}$, 
which is independent of the hypotheses being tested. 
The probability density $T_{ijk}$ is a normalized function of $\mathcal{D}_{ij}$ given a specific value of $m_{4\PGm}$,
obtained from simulation and including systematic variations through alternative distributions, as described below.
We assume the same quantum numbers and couplings for all signal resonances, enabling the use of a shared $T_{ijk}$
for parameterizing the signal.

To distinguish between alternative models, the test statistic $q=-2{\ln(\mathcal{L}_{J_j^P}/\mathcal{L}_{J_i^P})}$ 
is defined using the ratio of signal plus background likelihoods for the two signal hypotheses. 
The likelihood is maximized with respect to the nuisance parameters, which include the yields
of signal and background (bkg) processes and constrained parameters describing the systematic uncertainties.
To quantify the consistency of the observed test statistic $q_\text{obs}$ with the model $J_i^P$, 
the probability $p = P( q \leq q_\text{obs} \, | \, J_i^P +\text{bkg} )$ is determined under the signal-plus-background  
hypothesis using pseudoexperiments. This probability is then translated into a $Z$-score, representing the number 
of standard deviations via the one-sided Gaussian tail integral.

The consistency of the observed data with the alternative signal hypothesis ($J_j^P$) is assessed from
$P( q \geq q_\text{obs} \, | \, J_j^P + \text{bkg} )$. 
The sign is positive if the tail extends away or negative if it extends towards the median of the other hypothesis.
The \CLs~criterion~\cite{Junk:1999kv,0954-3899-28-10-313}, defined
as $\CLs = { P( q \geq q_\text{obs} \, | \, J_j^P + \text{bkg} ) }$ / ${ P( q \geq q_\text{obs} \, | \, J_i^P + \text{bkg} ) } < \alpha,$
is used for the final inference of whether a particular alternative hypothesis $J_j^P$ is excluded or not 
with respect to a reference hypothesis $J_i^P$ at a given confidence level $(1 -\alpha)$.
Figure~\ref{fig:Dij} (right) shows example $q$ distributions for the $2_m^{+}$ and $0^{-}$ models,
obtained from repeated pseudoexperiments simulating the expected experimental outcome. 

\begin{table*}[t!]
  \centering
      \topcaption{
      \textbf{Results from hypothesis test for pairs of spin-parity models.} 
      This is an extended version of Table~\ref{table:results-short}.
      The expected $p$-value is presented based on the assumption of the scenario of $2_m^+$.
      Results with $Z>5$ have been derived through Gaussian extrapolation.}
      \begin{tabular}{llcccc}
      \hline
       & & \multicolumn{2}{c}{Observed}  & \multicolumn{2}{c}{Expected} \\
      &&  $p$-value & $Z$-score & $p$-value & $Z$-score \\
      \hline
      \multirow{2}{*}{$0^{-}$ vs $2^{+}_m$} & $0^{-}$ & $2.7 \times 10^{-13}$ & $7.2$ & $6.5 \times 10^{-14}$ & $7.4$ \\
      & $2^{+}_m$ & $0.42$ & $0.2$ & $0.5$ & $0$ \\[\cmsTabSkip]
      \multirow{2}{*}{$0_{m}^{+}$ vs $2^{+}_m$} & $0_{m}^{+}$ & $4.3 \times 10^{-5}$ & $3.9$ & $5.6 \times 10^{-9}$ & $5.7$ \\
      & $2^{+}_m$ & $0.072$ & $1.5$ & $0.5$ & $0$ \\
      \multirow{2}{*}{$0_{\text{mix}}^{+}$ vs $2^{+}_m$} & $0_{\text{mix}}^{+}$ & $1.4 \times 10^{-2}$ & $2.2$ & $8.4 \times 10^{-4}$ & $3.1$ \\
      & $2^{+}_m$ & $0.17$ & $1.0$ & $0.5$ & $0$ \\
      \multirow{2}{*}{$0_{h}^{+}$ vs $2^{+}_m$} & $0_{h}^{+}$ & $3.1 \times 10^{-9}$ & $5.8$ & $8.5 \times 10^{-5}$ & $3.8$ \\
      & $2^{+}_m$ & $0.90$ & $-1.3$ & $0.5$ & $0$ \\[\cmsTabSkip]
      \multirow{2}{*}{$1^{-}$ vs $2^{+}_m$} & $1^{-}$ & $8.0 \times 10^{-8}$ & $5.2$ & $6.4 \times 10^{-9}$ & $5.7$ \\
      & $2^{+}_m$ & $0.38$ & $0.3$ & $0.5$ & $0$ \\[\cmsTabSkip]
      \multirow{2}{*}{$1^{+}$ vs $2^{+}_m$} & $1^{+}$ & $4.7 \times 10^{-3}$ & $2.6$ & $2.7 \times 10^{-5}$ & $4.0$ \\
      & $2^{+}_m$ & $0.052$ & $1.6$ & $0.5$ & $0$ \\[\cmsTabSkip]
      \multirow{2}{*}{$2_{m}^{-}$ vs $2^{+}_m$} & $2_{m}^{-}$ & $4.1 \times 10^{-12}$ & $6.8$ & $3.9 \times 10^{-14}$ & $7.5$ \\
      & $2^{+}_m$ & $0.28$ & $0.6$ & $0.5$ & $0$ \\
      \multirow{2}{*}{$2_{\text{mix}}^{-}$ vs $2^{+}_m$} & $2_{\text{mix}}^{-}$ & $6.5 \times 10^{-4}$ & $3.2$ & $1.5 \times 10^{-4}$ & $3.6$ \\
      & $2^{+}_m$ & $0.31$ & $0.5$ & $0.5$ & $0$ \\
      \multirow{2}{*}{$2_{h}^{-}$ vs $2^{+}_m$} & $2_{h}^{-}$ & $2.2 \times 10^{-8}$ & $5.5$ & $6.3 \times 10^{-9}$ & $5.7$ \\
      & $2^{+}_m$ & $0.43$ & $0.2$ & $0.5$ & $0$ \\
      \hline
      \end{tabular}
      \label{table:results}
  \end{table*}

The pairs of spin-parity models are tested among the 
$0^-$, $0_m^+$, $0_h^+$, $1^-$, $1^+$, $2_m^-$, $2_h^-$, and $2^{+}_m$ hypotheses. 
In all tests involving the $2^{+}_m$ model, it is preferred.
Therefore, these tests are presented in Fig.~\ref{fig:summary} and Table~\ref{table:results},
which is an extended version of Table~\ref{table:results-short}.
The pairwise tests between the other models do not provide any additional useful information, 
as the data frequently show inconsistencies with both models.
In the case of $J^{PC}=0^{++}$ and $2^{-+}$ scenarios, additional tests are conducted to account for 
a possible mixture of two tensor structures, as outlined in Section~\ref{sec:nature_results}.

It is important to note that the $2^{+}_m$ model represents only one specific realization of the $J^{PC} = 2^{++}$ 
hypothesis. In practice, a  mixture of amplitudes corresponding to the $2^{++}$ state, as listed in Table~\ref{table:jcp}, 
could produce angular distributions that resemble those expected for $0^{++}$ or $1^{++}$, in addition to $2^{+}_m$. 
As a result, even if the true particle is a $2^{++}$ state, the $p$-values reported in Table~\ref{table:results} may 
not remain fully consistent with the $2^{+}_m$ model under such admixtures.
These mixed scenarios can be explored in future analyses.
Taking this into account, all observed data distributions are found to be compatible with the 
$J^{PC} = 2^{++}$ hypothesis and show deviations from the predictions of alternative $J^{PC}$ 
assignments, with confidence levels summarized in Table~\ref{table:results}.
\cleardoublepage \section{The CMS Collaboration \label{app:collab}}\begin{sloppypar}\hyphenpenalty=5000\widowpenalty=500\clubpenalty=5000\cmsinstitute{Yerevan Physics Institute, Yerevan, Armenia}
{\tolerance=6000
A.~Hayrapetyan, V.~Makarenko\cmsorcid{0000-0002-8406-8605}, A.~Tumasyan\cmsAuthorMark{1}\cmsorcid{0009-0000-0684-6742}
\par}
\cmsinstitute{Institut f\"{u}r Hochenergiephysik, Vienna, Austria}
{\tolerance=6000
W.~Adam\cmsorcid{0000-0001-9099-4341}, J.W.~Andrejkovic, L.~Benato\cmsorcid{0000-0001-5135-7489}, T.~Bergauer\cmsorcid{0000-0002-5786-0293}, M.~Dragicevic\cmsorcid{0000-0003-1967-6783}, C.~Giordano, P.S.~Hussain\cmsorcid{0000-0002-4825-5278}, M.~Jeitler\cmsAuthorMark{2}\cmsorcid{0000-0002-5141-9560}, N.~Krammer\cmsorcid{0000-0002-0548-0985}, A.~Li\cmsorcid{0000-0002-4547-116X}, D.~Liko\cmsorcid{0000-0002-3380-473X}, M.~Matthewman, I.~Mikulec\cmsorcid{0000-0003-0385-2746}, J.~Schieck\cmsAuthorMark{2}\cmsorcid{0000-0002-1058-8093}, R.~Sch\"{o}fbeck\cmsAuthorMark{2}\cmsorcid{0000-0002-2332-8784}, D.~Schwarz\cmsorcid{0000-0002-3821-7331}, M.~Shooshtari, M.~Sonawane\cmsorcid{0000-0003-0510-7010}, W.~Waltenberger\cmsorcid{0000-0002-6215-7228}, C.-E.~Wulz\cmsAuthorMark{2}\cmsorcid{0000-0001-9226-5812}
% ...
\par}
\vskip\cmsinstskip
\dag:~Deceased\\
$^{1}$Also at Yerevan State University, Yerevan, Armenia\\
$^{2}$Also at TU Wien, Vienna, Austria\\
% ...
\end{sloppypar}
\end{document}